\documentclass[12pt]{article}
\usepackage{amsfonts}
\usepackage{amsmath}
\usepackage{pdfsync}
\usepackage{amsmath}
\usepackage{amsfonts,amssymb}
\usepackage{color}

\usepackage[english]{babel}
\usepackage[latin1]{inputenc}
\usepackage[T1]{fontenc}

\newtheorem{theorem}{Theorem}
\newtheorem{definition}{Definition}
\newtheorem{lemma}{Lemma}
\newtheorem{remark}{Remark}
\newtheorem{corollary}{Corollary}
\newtheorem{proposition}{Proposition}

\numberwithin{equation}{section}

\begin{document}

\title{\bf Quasi-invariant states }
\date{}
\maketitle
\author{}

\centerline{{Luigi Accardi}}

\centerline{\small{Volterra Center, University of Roma Tor Vergata}}
\centerline{\small{Via Columbia 2, 00133 Roma, Italy}}
\centerline{\small{accardi@volterra.uniroma2.it}}
\centerline{{Ameur Dhahri}}
\centerline{\small{Dipartimento di Matematica, Politecnico di Milano}}
\centerline{\small{Piazza Leonardo da Vinci 32, I-20133 Milano, Italy}}
\centerline{\small{ameur.dhahri@polimi.it}}

\begin{abstract}
We develop the theory of quasi--invariant (resp. strongly quasi--invariant) states under the action of a group $G$ of normal $*$--automorphisms of a $*$--algebra (or von Neumann alegbra) $\mathcal{A}$.
We prove that these states are naturally associated to left--$G$--$1$--cocycles.
If $G$ is compact, the structure of strongly $G$--quasi--invariant states is determined.
For any $G$--strongly quasi--invariant state $\varphi$, we construct a unitary representation
associated to the triple $(\mathcal{A},G,\varphi)$.
We prove, under some conditions, that any quantum Markov chain with commuting, invertible and hermitean conditional
density amplitudes on a countable tensor product of type I factors is strongly quasi--invariant with respect to the natural action of the group $\mathcal{S}_{\infty}$
of local permutations and we give the explicit form of the associated cocycle.
This provides a family of non--trivial examples of strongly quasi--invariant states for
locally compact groups obtained as inductive limit of an increasing sequence of compact
groups.
\end{abstract}

\tableofcontents

\section{Introduction}


Given a faithful state on $*$--algebra $\mathcal{A}$ and a group $G$ of
$*$-automorphisms of $\mathcal{A}$, we say that $\varphi$ is $G$--invariant if
\begin{equation}\label{inv}
\varphi(g(a))=\varphi(a),\quad \forall g\in G,\;\;a\in\mathcal{A}
\end{equation}
This class of states is of paramount importance in many branches of mathematics and physics and
it has been studied in different contexts (cf \cite{[BratRobins]}, \cite{[CD78]}, \cite{[HerTakes]}, \cite{[HLS81]}, \cite{[Stormer]}, \cite{[Stormer71]}...). In this paper, 
we extend definition \eqref{inv} as follows: we assume that for every
$g\in G$ there exists $x_g\in\mathcal{A}$ such that
\begin{equation}\label{q-inv}
\varphi(g(x))=\varphi(x_ga)
\end{equation}
and we call $G$--quasi--invariant a state satisfying \eqref{q-inv}.
Moreover, if in addition $x_g=x_g^*$ for every $g\in G$, the state $\varphi$ is said to be $G$--strongly quasi--invariant. In this case we prove that the Radon--Nikodym derivatives $x_g$'s
are invertible positive operators and satisfy the normalized left--$G$--$1$--cocycle property
$$x_{g_{2}g_{1}}  =  x_{g_{1}}(g_{1}^{-1}(x_{g_{2}})),\quad x_e=1$$
Moreover, we show that the $*$--algebra generated by the Radon--Nikodym derivatives $x_g$'s is commutative. We further study the case of a  compact group $G$ of normal $*$--automorphisms of a von Neumann algebra $\mathcal{A}$. In particular, we give the structure of a normal $G$--strongly quasi--invariant state $\varphi$, when $G$ is a compact group, and we prove that it has the following form
\begin{equation}\label{struc}
\varphi(a)=\varphi_G(\kappa_G^{-1}a), \quad \forall a\in \mathcal{A},
\end{equation}
for some bounded invertible positive operator $\kappa_G$ in the centralizer of $\varphi$ whose structure is explicitly given (see Theorem \ref{thm:compact-case}). Conversely, if $\varphi_G$ is a $G$--invariant state and $\kappa_G$ is an invertible positive operator in the centralizer of $\varphi$, then the state $\varphi$, defined by the right hand side of \eqref{struc}, is a $G$--quasi--invariant state with cocycle
\begin{equation}\label{struc-1cocy}
x_g=\kappa_G g^{-1}(\kappa_G^{-1}), \quad \forall g\in G
\end{equation}
We study the inductive limit associated to a group $G$ which is the union of an increasing sequence of compact groups. For any $G$--strongly quasi--invariant state  $\varphi$, the unitary representation of the group $G$ is also given.
Moreover, we give some examples and we investigate the structure of a
$\mathcal{S}_\infty$--quasi--invariant state $\varphi$, where
$\mathcal{S}_\infty=\cup_{N=1}^\infty\mathcal{S}_N$ is the group of local permutations on an infinite tensor product of a von Neumann algebra ($\mathcal{S}_N$ is the group of permutations on $\{1,2,\dots,N\}$).



\section{Quasi--invariant states under a group of $*$-automorphisms}\label{sec:G-QIS}

In the following $\mathcal{A}$ will denote a $*$--algebra, $\varphi$ a
\textbf{faithful} state on $\mathcal{A}$ and $G\subseteq\hbox{Aut}(\mathcal{A})$ a
group of $*$--automorphisms of $\mathcal{A}$.
\begin{proposition}\label{prop:xg-1G-left-cocy}{\rm
Suppose that, for any $g\in G$, there exists a $x_{g}\in\mathcal{A}$ such that
\begin{equation}\label{df-G-q-inv}
\varphi(g(a))=\varphi(x_{g}a)\ ;\qquad\forall\,a\in\mathcal{A}
\end{equation}
Then the map $g\in G\mapsto x_{g}\in\mathcal{A}$
is a \textbf{normalized multiplicative left $G$--$1$--cocycle}, i.e. it satisfies the identities
\begin{equation}\label{xe=1}
x_{e} = 1
\end{equation}
\begin{equation}\label{xg-1G-left-cocy}
x_{g_{2}g_{1}}  = x_{g_{1}} g_{1}^{-1}(x_{g_{2}})
\quad,\quad\forall g_{2}, g_{1}\in G
\end{equation}
In particular each $x_{g}$ is \textbf{invertible} and its inverse is
\begin{equation}\label{xg-1}
x_{g}^{-1} =   g^{-1}(x_{g^{-1}})
\left(\iff  x_{g^{-1}} = g(x_{g}^{-1})  \right)
\end{equation}
}\end{proposition}
\textbf{Proof}. One has
$$
\varphi(a)=\varphi(e(a))=\varphi(x_{e}a)\ ;\qquad\forall\,a\in\mathcal{A}
$$
Since $\varphi$ is faithful, \eqref{xe=1} follows.
Now for any $g_{2}, g_{1}\in G$, one has
$$
\varphi(g_{2}g_{1}(a))=\varphi(x_{g_{2}g_{1}}a)\\
=\varphi(x_{g_{2}}(g_{1}(a)))\\
=\varphi(g_{1} (g_{1}^{-1}(x_{g_{2}})a) )\\
$$
$$
=\varphi(x_{g_{1}}(g_{1}^{-1}(x_{g_{2}})a) )\\
=\varphi((x_{g_{1}}g_{1}^{-1}(x_{g_{2}}))a )
$$
This is equivalent to
\begin{eqnarray*}
 0&=& \varphi(x_{g_{2}g_{1}}a) -  \varphi(x_{g_{1}}(g_{1}^{-1}(x_{g_{2}}))a )\\
&=&\varphi( (x_{g_{2}g_{1}}-x_{g_{1}}(g_{1}^{-1}(x_{g_{2}})))a)
\end{eqnarray*}
Since $a$ is arbitrary, one can choose $a= (x_{g_{2}g_{1}}-x_{g_{1}}(g_{1}^{-1}(x_{g_{2}})))^*$.
This gives
$$
0=\varphi( |(x_{g_{2}g_{1}}-x_{g_{1}}(g_{1}^{-1}(x_{g_{2}})))^*|^2)
$$
Since $\varphi$ is faithful, \eqref{xg-1G-left-cocy} follows.
Taking $g_{2} := g_{1}^{-1}$ in \eqref{xg-1G-left-cocy}, one finds, using $x_{e} = 1$,
$$
1 = x_{e}  =  x_{g_{1}}(g_{1}^{-1}(x_{g_{1}^{-1}}))
$$
This implies that $x_{g_{1}}$ is invertible and its inverse is
$$
x_{g_{1}}^{-1} =   g_{1}^{-1}(x_{g_{1}^{-1}})
$$
since $g_{1}$ is arbitrary, the first identity in \eqref{xg-1} follows. Multiplying both sides of it
by $g_{1}$, one obtains
$$
g_{1}\left( x_{g_{1}}^{-1}\right) =   x_{g_{1}^{-1}}
$$
which is equivalent to the second identity in \eqref{xg-1}.
$\qquad\square$
\\
The following result is known. We include a simple proof of it for completeness.
\begin{lemma}\label{g-aut-fg=gf}{\rm
If $\mathcal{A}$ is a von Neumann algebra, $g$ is a normal $*$--automorphism of $\mathcal{A}$ and if
$$
x=g(y)
$$
with $x$ and $y$ are selfadjoint, invertible in $\mathcal{A}$. Then
\begin{equation}\label{xs=g(ys)}
x^{s}=g(y^{s})\quad,\quad \forall s\in\mathbb{R}
\end{equation}
}\end{lemma}
\textbf{Proof}.
$$
1=g(yy^{-1})=g(y)g(y^{-1})
$$
Therefore $g(y)$ is invertible and
\begin{equation}\label{inv1}
g(y)^{-1}=g(y^{-1})=x^{-1}
\end{equation}
Thus
$$
x^{s}=g(y^{s})\quad,\quad\forall s\in\mathbb{Z}
$$
For $n\in\mathbb{N}$ one has
$$
g(y^{\frac{1}{n}})^{n} = g((y^{\frac{1}{n}})^{n})=g(y)=x
$$
Therefore
$$
x^{\frac{1}{n}}=g(y^{\frac{1}{n}})\quad;\quad \forall n\in\mathbb{N}
$$
Consequently
$$
x^{\frac{m}{n}}=(x^{\frac{1}{n}})^{m}=g(y^{\frac{1}{n}})^{m}
=g(y^{\frac{m}{n}})
$$
and, because of \eqref{inv1}, $x^{s}=g(y^{s})$ for all $s\in \mathbb{Q}$.
By continuity
$$
x^{s}=g(y^{s})\quad,\quad\forall s\in\mathbb{R}
$$
$\square$
\begin{corollary}{\rm
Under the assumptions of Lemma \ref{g-aut-fg=gf}, if $x_g=x_g^*$, then for any $s\in\mathbb{R}$,
\begin{equation}\label{x(-s)g}
x^{-s}_g=g^{-1}(x^s_{g^{-1}})
\end{equation}
}\end{corollary}
\textbf{Proof}.
Replacing in \eqref{xs=g(ys)} $g$ by $g^{-1},x$ by $x_{g}^{-1}$ and $y$ by $x_{g^{-1}}$, one finds
$$
x_{g}^{-s}=g^{-1}(x_{g^{-1}}^{s})
$$
which is \eqref{x(-s)g}.
$\square$
\begin{definition}\label{df:G-q-inv}{\rm
If the pair $(G,x)$ satisfies condition \eqref{df-G-q-inv} of Proposition \ref{prop:xg-1G-left-cocy}
the state $\varphi$ is called $(G,x)$--\textbf{quasi--invariant}, or $G$--quasi--invariant with cocycle $x$.
}\end{definition}
\begin{lemma}{\rm
If $\varphi$ is $(G,x)$--quasi--invariant, then for all $g\in G$
\begin{equation}\label{xg-normaliz}
a\in\mathcal{A} \ , \ a\ge 0 \Rightarrow \varphi(x_{g}a)\ge 0
 \quad ,\quad  1 =\varphi(x_{g})
\end{equation}
\begin{equation}\label{almst-ctrliz}
\varphi(x_{g}a)=\varphi(ax_{g}^*)
\ ;\qquad\forall\,a\in\mathcal{A}
\end{equation}
}\end{lemma}
\textbf{Proof}. If $a\in\mathcal{A} \ , \ a\ge 0$ and $g\in G$, one has
$0\le\varphi(g(a))=\varphi(x_{g}a)$ and this proves the inequality in \eqref{xg-normaliz}.
The equality follows from
$$
1=\varphi(1)=\varphi(g(1))=\varphi(x_{g})\ ;\qquad\forall\,g\in G
$$
Moreover one has
$$
\varphi(g(a^*))=\varphi(x_{g}a^*)\ ;\qquad\forall\,a\in\mathcal{A}
$$
On the other hand
$$
\varphi(g(a^*))
=\varphi((g(a))^*)
=\overline{\varphi(g(a))}
=\overline{\varphi(x_{g}a)}
=\varphi((x_{g}a)^*)
=\varphi(a^*x_{g}^*)
$$
It follows that
$$
\varphi(x_{g}a^*)=\varphi(a^*x_{g}^*)   \ ;\qquad\forall\,a\in\mathcal{A}
$$
and, since $a\in\mathcal{A}$ is arbitrary, this is equivalent to \eqref{almst-ctrliz}.
$\qquad\square$
\begin{remark}{\rm
Recall that the \textbf{ centralizer\/} of $\varphi$, hereinafter denoted
$\hbox{Centrz}(\varphi)$, is characterized by the property
\begin{equation}\label{df-centrliz-varphi}
c\in\ \hbox{Centrz}(\varphi)\Leftrightarrow\varphi(ac)=\varphi(ca)\quad ;\quad
\forall\,a\in\mathcal{A}
\end{equation}
}\end{remark}
\begin{lemma}\label{lm:xg-herm}{\rm
If $\varphi$ is $(G,x)$--quasi--invariant, for any $g\in G$ the following are equivalent:
\begin{enumerate}
\item[(i)] $x_{g}$ is a hermitean element of $\mathcal{A}$
\item[(ii)] $x_{g}$ is in the centralizer of $\varphi$.
\end{enumerate}
}\end{lemma}
\textbf{Proof}.
(i) $\Rightarrow$ (ii). If $x_{g}= x_{g}^*$, (ii) follows from \eqref{almst-ctrliz}.\\
(ii) $\Rightarrow$ (i). If $x_{g}$ is in the centralizer of $\varphi$ then the same is true for
$x_{g}^*$. Therefore \eqref{almst-ctrliz} implies that for all $a\in\mathcal{A}$,
$$
\varphi(x_{g} a)
=\varphi(ax_{g}^*)
=\varphi(x_{g}^*a)
\iff \varphi((x_{g}-x_{g}^*) a)=0
\overset{a:=(x_{g}-x_{g}^*)^*}{\Rightarrow} \ \varphi(|x_{g}-x_{g}^*|^2)=0
$$
and (i) follows because $\varphi$ is faithful.
$\square$



\begin{definition}\label{def-strong}{\rm
A $(G,x)$--quasi--invariant state $\varphi$ is called $(G,x)$--\textbf{strongly quasi--invariant}
if, for any $g\in G$, $x_{g}$ is \textbf{hermitean}.
}\end{definition}

\begin{lemma}\label{lm:the-xg-commute}{\rm
If $\varphi$ is $(G,x)$--strongly quasi--invariant, $x$ satisfies
\begin{equation}\label{xg-comm-g(xg')}
x_{g_{1}}(g_{1}^{-1}(x_{g_{2}})) = (g_{1}^{-1}(x_{g_{2}}))x_{g_{1}}
\quad,\quad \forall g_{2}, g_{1}\in G
\end{equation}
which is equivalent to
\begin{equation}\label{xg-comm-xg'}
x_{g_{2}}x_{g_{1}} = x_{g_{1}}x_{g_{2}}
\quad,\quad \forall g_{2}, g_{1}\in G
\end{equation}
Moreover for all $g\in G$, $x_{g}$ is an invertible positive operator.
}\end{lemma}
\textbf{Proof}.
From the cocycle identity (\ref{xg-1G-left-cocy}) and the hermiteanity of the $x_{g}$, one has,
for all $g_{2}, g_{1}\in G$,
$$
x_{g_{1}}(g_{1}^{-1}(x_{g_{2}})) = x_{g_{2}g_{1}} =  x_{g_{2}g_{1}}^*
=  (g_{1}^{-1}(x_{g_{2}}))x_{g_{1}}
$$
which is \eqref{xg-comm-g(xg')}. \eqref{xg-comm-g(xg')} is equivalent to
$$
(g_{1}^{-1}(x_{g_{2}})) x_{g_{1}}^{-1}
=  x_{g_{1}}^{-1}(g_{1}^{-1}(x_{g_{2}}))
\overset{\eqref{xg-1}}{\iff}
(g_{1}^{-1}(x_{g_{2}}))g_{1}^{-1}(x_{g_{1}})
= g_{1}^{-1}(x_{g_{1}})(g_{1}^{-1}(x_{g_{2}}))
$$
$$
\iff  g_{1}^{-1}(x_{g_{2}}x_{g_{1}}) = g_{1}^{-1}(x_{g_{1}}x_{g_{2}})
\iff  x_{g_{2}}x_{g_{1}} = x_{g_{1}}x_{g_{2}}
$$
which is \eqref{xg-comm-xg'}.

Note that $x_{g}=x_{g}^*=x_{g,+}-x_{g,-}$, with $x_{g,\pm}$ are positive such $x_{g,+}x_{g,-}=0$.
Therefore, choosing in \eqref{almst-ctrliz} $a=x_{g,-}$, one finds
$$
0\le \varphi(x_{g}x_{g,-})=\varphi((x_{g,+}-x_{g,-})x_{g,-})=-\varphi(x_{g,-}^2)
\iff 0=\varphi(x_{g,-}^2) \iff 0=x_{g,-}
$$
Thus $x_{g}=x_{g}^*=x_{g,+}\ge 0$ and by Proposition \ref{prop:xg-1G-left-cocy},
$x_g$ is invertible.
$\square$

Let $\mathcal{C}$ be the $*$--algebra generated by the $x_{g}\;(g\in G)$. It follows from Lemma \ref{lm:xg-herm} that $\mathcal{C}\subseteq \hbox{Centrz}(\varphi)$ and $\mathcal{C}$ is  abelian by Lemma \ref{lm:the-xg-commute}.
\begin{lemma}\label{lm:Centrz}{\rm
If $\varphi$ is $(G,x)$--strongly quasi--invariant
then for all $g\in G$ and $a\in\mathcal{A}$
\begin{equation}\label{varphi(g(x)a)}
\varphi(g(x)a)
=\varphi(ag(x_{g}xx_{g}^{-1}))\quad \forall x\in \hbox{Centrz}(\varphi)
\end{equation}
In particular, for all $g\in G$
\begin{equation}\label{g(C)-in-Centr(fi)}
g(\mathcal{C}) = \mathcal{C}\subseteq \hbox{Centrz}(\varphi)  \qquad,\quad\forall g\in G
\end{equation}
}\end{lemma}
\textbf{Proof}.
In the above notations, one has
$$
\varphi(g(x)a)=\varphi(g(xg^{-1}(a)))
=\varphi(x_{g}xg^{-1}(a))
$$
$$
=\varphi(g^{-1}(a)x_{g}x)
=\varphi(g^{-1}(ag(x_{g}x)))
=\varphi(x_{g^{-1}}ag(x_{g})g(x))
=\varphi(ag(x_{g})g(x)x_{g^{-1}})
$$
$$
\overset{\eqref{xg-1}}{=}\varphi(ag(x_{g})g(x)g(x_{g}^{-1}))
=\varphi(ag(x_{g}xx_{g}^{-1}))
$$
which is \eqref{varphi(g(x)a)}. Now if $x\in \mathcal{C}\subset\hbox{Centrz}(\varphi)$ and if $g\in G$, one has $x_{g}xx_{g}^{-1}=x$. This proves the inclusion in
\eqref{g(C)-in-Centr(fi)}. The identity in \eqref{g(C)-in-Centr(fi)} follows from
$$
x_{g_{2}g_{1}}  \overset{\eqref{xg-1G-left-cocy}}{=} x_{g_{1}} g_{1}^{-1}(x_{g_{2}})
\iff x_{g_{1}}^{-1}x_{g_{2}g_{1}}  =  g_{1}(x_{g_{2}})
\iff  g_{1}(x_{g_{2}})\in\mathcal{C}
$$
because $g_{1}$ and $x_{g_{2}}$ are arbitrary and $g_{1}$ is invertible.
$\square$


\subsection{The case $\mathcal{A}=\mathcal{B}(\mathcal{H})$}

The following Proposition shows that, if $\mathcal{A}=\mathcal{B}(\mathcal{H})$,
there exist both\\ $G$--quasi--invariant states and $G$--strongly quasi--invariant states.
\begin{proposition}\label{prop:xh-BH-case}{\rm If $\mathcal{A}=\mathcal{B}(\mathcal{H})$ and $\varphi(.)=\mbox{Tr}(W \, \cdot \,) $ is a faithful state, where $W$ is
density operator, then  $x_{g}$ is a solution of equation \eqref{almst-ctrliz} if and only if
\begin{equation}\label{(Wxg=xg*W)}
Wx_{g}=x_{g}^{*}W
\end{equation}
and the invertible solutions of equation \eqref{(Wxg=xg*W)} are given by the set
\begin{equation}\label{df-SW}
S_{W}:=\{W^{-1}z \colon z\in\mathcal{B}(\mathcal{H})\quad,\quad z=z^{*}\}
\end{equation}
}\end{proposition}
\textbf{Proof}.
In the assumptions of the Proposition, equation \eqref{almst-ctrliz}, i.e.
\begin{equation}\label{almst-ctrliz2}
\varphi(x_{g}a)=\varphi(ax_{g}^*)   \ ;\qquad\forall\,a\in\mathcal{A}
\end{equation}
becomes equivalent to
$$
\mbox{Tr}(Wx_{g}a)=\mbox{Tr}(Wax_{g}^{*})
=\mbox{Tr}(x_{g}^{*}Wa)\quad ;\quad \forall a
$$
$$
\iff Wx_{g}=x_{g}^{*}W
$$
which is \eqref{(Wxg=xg*W)}.
If $x$ is a solution of equation \eqref{(Wxg=xg*W)}, defining
$$
z:= Wx
$$
one has
$$
z^{*}=(Wx)^{*}=x^{*}W\overset{\eqref{(Wxg=xg*W)}}{=}Wx=z
$$
thus $x=W^{-1}z\in S_{W}$. Conversely, if $x\in S_{W}$, then $x=W^{-1}z$ for some
$z=z^{*}\in\mathcal{B}(\mathcal{H})$.
Therefore
$$
Wx=W(W^{-1}z)=z=zW^{-1}W
=(W^{-1}z)^{*}W=x^{*}W
$$
i.e. $x$ is a solution of \eqref{(Wxg=xg*W)}.
$\square$\\
\begin{remark}{\rm
1) If $z=z^{*}\in\mathcal{B}(\mathcal{H})$ commutes with $W$, then it commutes with $W^{-1}$
and therefore, if $x:=W^{-1}z$, one has
$$
x^{*}=zW^{-1}=W^{-1}z=x
$$
In other words, equation \eqref{(Wxg=xg*W)} always admits hermitean solutions when
$\mathcal{A}=\mathcal{B}(\mathcal{H})$ and these solutions are in
$\mathcal{B}(\mathcal{H})$ whenever $W$ is bounded away from $0$.\smallskip\\
2) From Proposition \ref{prop:xh-BH-case}, one also deduces that the family of strong quasi--invariant states is strictly smaller than the family of quasi--invariant states. In fact, let
$$
\mathcal{G}=\{g_{0}^{n}: n\in\mathbb{N}\}\quad;\quad g^{0}_{0}:=e\neq g_{0}
$$
be a group of $*$--automorphism of $\mathcal{B}(\mathcal{H})$ with a single generator $g_{0}\neq e$.\\
Given a density operator $W$ bounded away from $0$, fix $z_{g_{0}}=z_{g_{0}}^*$ subject to the only condition
\begin{equation}\label{xg-ne-0}
[z_{g_{0}}, W]\neq 0
\end{equation}
(clearly, for any $W$ as above, there are many $z_{g_0}$ satisfying
\eqref{xg-ne-0}) and define
\begin{equation}\label{xg0}
x_{g_{0}}:=W^{-1}z_{g_{0}}\in\mathcal{B}(\mathcal{H})
\end{equation}
The identity
$$
x_{g_{0}^{n}}=x_{g_{0}}g_{0}^{-1}(x_{g_{0}})\cdots g_{0}^{-(n-1)}(x_{g_{0}})
\quad,\quad n\in\mathbb{N}
$$
easily verified by induction on the cocycle equation, shows that \eqref{xg0} uniquely
determines $x_{g_{0}^{n}}$ for any $n\in\mathbb{N}$, i.e. the cocycle
$x\colon\mathcal{G}\rightarrow \mathcal{A}$.
On the other hand condition \eqref{xg-ne-0} means that $W^{-1}z_{g_{0}}$ is not in the centralizer
of $\varphi =\hbox{Tr}(W \, \cdot \, )$. Hence it is not hermitian by Lemma
\ref{lm:xg-herm}.
}\end{remark}

\subsection{Quasi--invariant states with trivial cocycles}\label{sec:QIS-triv-cocy}

\begin{lemma}\label{Lm:triv-l1cocy-unimod-G}{\rm

For any operator $\kappa\in\mathcal{A}$ invertible in $\mathcal{A}$, the map
\begin{equation}\label{df-triv-l1cocy}
g\in G\mapsto x_{g} = \kappa g^{-1}\left(\kappa^{-1}\right)
\end{equation}
is a \textbf{normalized left  multiplicative
$G$--$1$--cocycle}.
}\end{lemma}
\textbf{Proof}.
Condition \eqref{xe=1} follows from
$$
x_{e} = \kappa \kappa^{-1} = 1
$$
Condition \eqref{xg-1G-left-cocy} follows from
$$
x_{g_{2}g_{1}}
= \kappa (g_{2}g_{1})^{-1}\left(\kappa^{-1}\right)
= \kappa g_{1}^{-1}g_{2}^{-1}\left(\kappa^{-1}\right)
= \kappa g_{1}^{-1}\left(\kappa^{-1}\left(\kappa g_{2}^{-1}\left(\kappa^{-1}\right)\right)\right)
$$
$$
= \kappa g_{1}^{-1}(\kappa^{-1})g_{1}^{-1}\left(\kappa g_{2}^{-1}\left(\kappa^{-1}\right)\right)
=  x_{g_{1}}g_{1}^{-1}(x_{g_{2}})    \quad,\quad\forall g_{2}, g_{1}\in G
$$
$\square$
\begin{definition}\label{df:triv-cocy}{\rm
Let $G$ be a group of $*$--automorphisms of $\mathcal{A}$.
A multiplicative left $G$--$1$--cocycle of the form \eqref{df-triv-l1cocy}
is called \textbf{trivial}.}
\end{definition}
\begin{theorem}\label{thm:triv-cocy}{\rm
Let $G$ be a group of $*$--automorphisms of $\mathcal{A}$ and let $\kappa\in\mathcal{A}$ be an invertible operator with inverse in $\mathcal{A}$. Then, for any $G$--invariant
state $\varphi_G$, the state
\begin{equation}\label{quasi}
\varphi ( \, \cdot \, ) := \varphi_G (\kappa \, \cdot \, )
\end{equation}
is $(G,x)$--quasi-invariant with cocycle
\begin{equation}\label{cocy-kappa0}
x_g:=\kappa^{-1} g^{-1}(\kappa) \quad,\quad\forall g\in G
\end{equation}
$\varphi$ is strongly $(G,x)$--quasi-invariant with cocycle \eqref{cocy-kappa0} iff
\begin{equation}\label{varphi-G(kappa.)str-QI}
\kappa = \kappa^* \quad;\quad
\kappa g(\kappa) = g(\kappa)\kappa \ ,  \ \forall g\in G
\end{equation}
}
\end{theorem}
\textbf{Proof}.
The first statement follows from the fact that, if $\varphi_G$, $\varphi$
and $\kappa^{-1}$ are as in the statement, then for any $a\in\mathcal{A}$ and $g\in G$, one has
$$
\varphi(g(a))=\varphi_G(\kappa g(a))
=\varphi_G(g[g^{-1}(\kappa)a])
=\varphi_G([g^{-1}(\kappa)a])
$$
$$
=\varphi_G(\kappa \kappa^{-1} [g^{-1}(\kappa)a])
=\varphi((\kappa^{-1} g^{-1}(\kappa)a)
=\varphi(x_ga)
$$
The first identity in \eqref{varphi-G(kappa.)str-QI} follows by taking adjoints
of both sides of \eqref{quasi} as in the proof of Lemma \ref{lm:xg-herm}. Given the validity of the
first identity, the right hand side of \eqref{cocy-kappa0} is Hermitian iff, for all $g\in G$,
$$
\kappa^{-1} g^{-1}(\kappa)
= (\kappa^{-1} g^{-1}(\kappa)^*
=  g^{-1}(\kappa)\kappa^{-1}
$$
and, since $g^{-1}$ is arbitrary, this is equivalent to the second identity in \eqref{varphi-G(kappa.)str-QI}.
$\square$

\section{Examples of quasi--invariant states under permutations}\label{sec:Expls-str-qis-perm}

In this section, we construct examples of quasi--invariant states on infinite
tensor products of copies of a type I factor with the natural action given by the permutation
group on $\mathbb{N}^*$.
We show that, in this framework, examples of non--trivial cocycles naturally arise.

Let $\mathcal{B}:=\mathcal{B}(\mathcal{H})$ for some (separable) Hilbert space $\mathcal{H}$. Below the symbol "$\otimes$" will always refer to the minimal C*-tensor product.
Denote
\begin{equation}\label{alg-inf-tens-prod}
\mathcal{A}=\bigotimes_{\mathbb{N}^*}\mathcal{B}
\end{equation}
the tensor product of countably many copies of $\mathcal{B}$ ($\bigotimes_{\mathbb{N}^*}\mathcal{B}$ is the inductive limit of $\bigotimes_{m=1}^n\mathcal{B}$)
and let $j_n$ be the natural embedding of $\mathcal{B}$ onto the $n$--th factor of $\mathcal{A}$
(we also use the notation $j_n(\mathcal{B}) =:\mathcal{A}_n \equiv
\mathcal{B}\otimes\bigotimes_{\mathbb{N}^*\setminus \{n\}}1_{\mathcal{B}}$).
With these notations, introducing the \textit{local algebras}
\begin{equation}\label{notat-jn}
\mathcal{A}_{[M,N]} := \bigvee_{n\in [M,N]} \mathcal{A}_n
:= \hbox{algebra generated by the } j_n(\mathcal{B})=\mathcal{A}_n \, \colon \, n\in [M,N]
\end{equation}
$(M\le N \in\mathbb{N}$) whose elements are called \textit{local elements}, one has
\begin{equation}\label{alg-inf-tens-prod}
\mathcal{A}=\bigcup_{N\in\mathbb{N}^*}\mathcal{A}_{[1,N]}
\end{equation} 
Denote by $\mathcal{S}_N$ the permutation group on $\{1,\dots, N\}$. Define $\mathcal{S}_\infty=\bigcup_{N\in\mathbb N^*}\mathcal{S}_N$.
$\mathcal{S}_\infty$ has a natural action on $\mathcal{A}$ by $*$--automorphisms given by
$g\circ j_n := j_{gn}$ where we use the same symbol for the actions of $\mathcal{S}_\infty$ on
$\mathbb N^*$ and on $\mathcal{A}$. With this identification, $\mathcal{S}_\infty$ is called the
group of local permutations on $\mathcal{A}$. Recall that, for each $g\in\mathcal{S}_\infty$,
the support of $g$ is defined by
$$
\Lambda_g := \hbox{supp}(g)
:= \{n\in\mathbb{N}^* \colon gn\neq n\}
=\hbox{supp}(g^{-1})
$$
Denoting
$$
\mathcal{S}_{N} := \{g\in\mathcal{S}_\infty \colon \hbox{supp}(g)\subseteq [1,N]\}
$$
one has
\begin{equation}\label{S-inf=union-SN}
\mathcal{S}_{\infty} = \bigcup_{N\in\mathbb{N}^*}\mathcal{S}_{N}
\end{equation}
In the literature, $\mathcal{S}_{\infty}$--invariant states (or weights) are also called \textit{exchangeable}
and we will we use both terms in what follows.

\subsection{Product states}\label{sec:Prod-sts}

In this section, we use the symbolic expressions $\rm{Tr}_{\mathbb{N}^*}$ and
$\prod_{n\in\mathbb{N}^*}j_n(W_{n}) $ (see formula \eqref{gen-pr-st} below).
It is known that there is no general way to give a meaning to $\rm{Tr}_{\mathbb{N}^*}$
as a (non identically infinite) weight on $\mathcal{A}$ or to $\prod_{n\in\mathbb{N}^*}j_n(W_{n}) $
as an element of a closure of $\mathcal{A}$ in some representation.
However the combined expression
\begin{equation}\label{formal-expr}
\rm{Tr}_{\mathbb{N}^*}\left(\prod_{n\in\mathbb{N}^*}j_n(W_{n}) \ \cdot \  \right)
\end{equation}
is well defined even at algebraic level by the left hand side of \eqref{gen-pr-st} and intuitively
highlights the formal analogy with \eqref{quasi} in which $\rm{Tr}_{\mathbb{N}^*}$ plays the role
of the state $\varphi_{G}$ and $\prod_{n\in\mathbb{N}^*}j_n(W_{n}) $ the role of $\kappa$.

In subection \ref{sec:QMC-comm-CDA}, we will see that this formal analogy can also be used for Markov states on $\mathcal{A}$. In both cases the continuous classical analogues of \eqref{formal-expr} are
widely used in quantum field theory (Feynman integral) and they also appear in the theory of
quantum flows.
Furthermore, once correctly defined by \eqref{gen-pr-st}, the expression \eqref{formal-expr}
can be used to obtain simplified proofs of several results. This statement is illustrated in
the proofs of Proposition \ref{prop:ex1-prod-sts} and Theorem \ref{thm:QI-MC}.

\begin{proposition} \label{prop:ex1-prod-sts}\rm{
Any product state on $\mathcal{A}$
\begin{equation}\label{gen-pr-st}
\varphi
=\bigotimes_{n\in \mathbb{N}^*}\varphi_{n}
=\bigotimes_{n\in\mathbb{N}^*}\rm{Tr}(W_{n} \, \cdot \, )
=:\rm{Tr}_{\mathbb{N}^*}\left(\prod_{n\in\mathbb{N}^*}j_n(W_{n}) \, \cdot \, \right)
\end{equation}
with $W_{n}$ invertible for each $n\in\mathbb{N}^*$ and satisfying the condition
\begin{equation}\label{Wn>cn>0}
\forall n\in\mathbb{N} \ , \ \exists c_{n}>0 \ \colon \ W_{n} \ge c_{n}
\end{equation}
is $\mathcal{S}_{\infty}$--quasi--invariant with cocycle
\begin{equation}\label{cocy-pr--st}
x_{g} :=
\left(\prod_{n\in \Lambda_g}j_n(W^{-1}_{n})\right)
g^{-1}\left(\prod_{n\in \Lambda_g}j_{n}(W_{n})\right)
=\prod_{n\in \Lambda_g}j_n(W^{-1}_{n}W_{g^{-1}n})
\quad,\quad\forall g\in \mathcal S_\infty
\end{equation}
and it is $\mathcal{S}_{\infty}$--strongly quasi--invariant iff the $W_n$ mutually commute.
}
\end{proposition}
\textbf{Proof}.
Let $g\in \mathcal{S}_{\infty}$ with support $\Lambda_g$. Note that $g(\Lambda_g)=\Lambda_g$. Therefore, for any $n\in\Lambda_g$ and $a\in\mathcal{A}$ one has
\begin{eqnarray*}
\varphi(g(a))&=&\rm{Tr}_{\mathbb{N}^*}\left(\prod_{n\in\mathbb{N}^*}j_n(W_{n}) g(a)\right)\\
&=&\rm{Tr}_{\mathbb{N}^*}\left(g\left(g^{-1}\left(\prod_{n\in\mathbb{N}^*}j_n(W_{n})\right) a\right)\right)\\
&=&\rm{Tr}_{\mathbb{N}^*}\left(g^{-1}\left(\prod_{n\in\mathbb{N}^*}j_n(W_{n})\right) a\right)\\
&=&\rm{Tr}_{\mathbb{N}^*}\left(\prod_{n\in\mathbb{N}^*}j_{g^{-1}n}(W_{n}) a\right)\\
&=&\rm{Tr}_{\mathbb{N}^*}\left(\left(\prod_{n\in \Lambda_g^c}j_n(W_{n})\right)
\left(\prod_{n\in \Lambda_g}j_{g^{-1}n}(W_{n})\right) a\right)\\
&=&\rm{Tr}_{\mathbb{N}^*}\left(\left(\prod_{n\in \mathbb{N}^*}j_n(W_{n})\right)
\left(\prod_{n\in \Lambda_g}j_n(W^{-1}_{n})\right)
\left(\prod_{n\in\Lambda_g}j_{g^{-1}n}(W_{n})\right)a\right)\\
&=&\varphi\left(\left(\prod_{n\in \Lambda_g}j_n(W^{-1}_{n})\right)\left(\prod_{n\in \Lambda_g}j_{g^{-1}n}(W_{n})\right)a\right)
=\varphi\left(x_{g}a\right)
\end{eqnarray*}
Condition \eqref{Wn>cn>0} guarantees that $x_{g}$ is bounded and the definition of
$\Lambda_g$ that $x_{g}$ is Hermitian. Therefore $\varphi$ is $\mathcal{S}_{\infty}$--quasi--invariant with cocycle given by \eqref{cocy-pr--st}.
$\qquad\square$\\

\noindent
\textbf{Remark}.
1) Notice that condition \eqref{Wn>cn>0} is very strong. In fact,
if $W_{n}=\sum_{n\in\mathbb{N}}w_{n}P_{n}$ is the spectral decomposition of $W_{n}$, due to the inequality
$$
1=\hbox{Tr}(W_{n}) = \sum_{n\in\mathbb{N}}w_{n}\hbox{Tr}(P_{n})\ge \sum_{n\in\mathbb{N}}w_{n}
$$
condition \eqref{Wn>cn>0} can be satisfied only if the spectrum of $W_{n}$ has finite cardinality.
\smallskip\\
2) Notice that the cocycle $x_{g}$ is non--trivial because it has the form
\begin{equation}\label{cocy-pr--st-non-triv}
x_{g} = \prod_{n\in \Lambda_g}j_n(W^{-1}_{n})g^{-1}\left(\prod_{n\in \Lambda_g}j_{n}(W_{n})\right)
=: \kappa_{g} g^{-1}\left(\kappa_{g}^{-1}\right)
\quad,\quad\forall g\in \mathcal S_\infty
\end{equation}
so \eqref{df-triv-l1cocy} is not satisfied because $\kappa_{g}$ depends on $g$. The only way to make
it formally independent on $g$ is to use the identity
$$
x_{g}
= \prod_{n\in \mathbb{N}^*}j_n(W^{-1}_{n})g^{-1}\left(\prod_{n\in\mathbb{N}^*}j_{n}(W_{n})\right)
= \prod_{n\in \mathbb{N}^*}j_n(W^{-1}_{n})\prod_{n\in\mathbb{N}^*}j_{g^{-1}(n)}(W_{n})
$$
$$
= \prod_{n\in \mathbb{N}^*}j_n(W^{-1}_{n})j_{g^{-1}(n)}(W_{n})
= \prod_{n\in \Lambda_g}j_n(W^{-1}_{n})j_{g^{-1}(n)}(W_{n})
$$
which is formally of the form \eqref{df-triv-l1cocy} with
$\kappa:=\prod_{n\in \mathbb{N}^*}j_n(W^{-1}_{n})$. But this is only a formal expression
because, in general, there is no reasonable representation $\pi$ of $\mathcal{A}$ in the
bounded operators on some Hilbert space $\mathcal{H}$ such that expressions like
$\prod_{n\in \mathbb{N}^*}j_n(W^{-1}_{n})$ or $\prod_{n\in \mathbb{N}^*}j_n(W_{n})$ can be
interpreted as operators in the closure of $\pi(\mathcal{A})$ for some topology on
$\mathcal{B}(\mathcal{H})$.\\
The above discussion suggests the following definition.
\begin{definition}\label{df:loc-triv-cocy}\upshape
Let $\mathcal{A}$, $(\mathcal{A}_{[1,N]})_{N\in\mathbb{N}^*}$ and $\mathcal{S}_{\infty}$ be as above.
An $\mathcal{S}_{\infty}$--cocycle $x\colon\mathcal{S}_{\infty}\to\mathcal{A}$ is called
\textit{locally trivial} if, denoting for $N\in\mathbb{N}^*$
$$
\mathcal{S}_{[1,N]}
:=\{g\in \mathcal{S}_{\infty} \colon g[1,N] = [1,N] \}
$$
there exists $\kappa_{[1,N]}\in \mathcal{A}_{[1,N]}$, with inverse in $\mathcal{A}_{[1,N]}$ such
that, for each $g\in\mathcal{S}_{[1,N]}$,
\begin{equation}\label{struct-loc-triv-cocy}
x_{g}
= \kappa_{[1,N]} g^{-1}\left(\kappa_{[1,N]}^{-1}\right)
\end{equation}
\end{definition}

\begin{theorem}\label{QI-of-prod-states}\upshape
Let $\varphi := \bigotimes_{\mathbb{N}^*} \hbox{Tr}(W_{k} \, \cdot \, )$ be an arbitrary
product state on $\mathcal{A}$ and let $W_{\infty}\in \mathcal{B}$ be a density operator bounded
away from zero. Denote
\begin{equation}\label{psi-hom-pr-st0}
\psi:=\bigotimes_{\mathbb{N}^*}\psi_{0}
=\bigotimes_{\mathbb{N}^*}\rm{Tr}_{\mathbb{N}^*}(W_{\infty} \, \cdot \, )
\end{equation}
the exchangeable state on $\mathcal{A}$ defined by $W_{\infty}$ and
$(\mathcal{H}_{\psi}, \pi_{\psi}, \Psi)$ the cyclic representation of $\psi$.
Then:\\
(i) The state $\hat{\psi}$ on $\pi_{\psi}(\mathcal{A})$ obtained by restriction on
$\pi_{\psi}(\mathcal{A})$ of the state
\begin{equation}\label{hat-psi}
\bar{a}\in\pi_{\psi}(\mathcal{A})''\mapsto \langle \Psi \, , \, \bar{a}\Psi\rangle
\quad,\quad  \bar{a}\in\pi_{\psi}(\mathcal{A})''
\end{equation}
is exchangeable, for the action of $\mathcal{S}_{\infty}$ on $\mathcal{A}$ defined by
$g\pi_{\psi}( \, \cdot \, ) := \pi_{\psi}(g( \, \cdot \, ))$ ($g\in\mathcal{S}_{\infty}$).\\
(ii) $\varphi$ is $\mathcal{S}_{\infty}$--quasi invariant with cocycle
\begin{equation}\label{cocycle-expl-prod-cse}
x_{g}:= \left(\prod_{n\in \Lambda_{g}}j_n(W_{\infty}^{-1}W_{n}^{-1})\right)
g^{-1}\left(\prod_{n\in \Lambda_{g}}j_{n}(W_{\infty}^{-1}W_{n})\right)
\in\mathcal{A}_{\Lambda_{g}}
\end{equation}
and it is $\mathcal{S}_{\infty}$--strongly quasi--invariant iff all $W_\infty$ and $W_n$ 's commute.\\
(iii) The weak limit
\begin{equation}\label{lim-x[0,n]0}
\lim_{n\to +\infty}\pi_{\psi}\left(\prod_{k\in [1,n]}j_k(W_{\infty}^{-1}W_{k})\right)
=: X_{\psi, \infty}
\end{equation}
exists in the strongly finite sense on the norm--dense subspace\\ $\bigcup_{n\in\mathbb{N}^*}\pi_{\psi}(\mathcal{A}_{[1,M]})''\cdot\Psi \subset \mathcal{H}_{\psi} $
and satisfies
\begin{equation}\label{varphi-hat-psi}
\varphi (a) = \hat{\psi}(X_{\psi, \infty}\pi_{\psi}(a))
\quad,\quad\forall a\in \mathcal{A}
\end{equation}
Moreover, $X_{\psi, \infty}\ge 0$ if all $W_{\infty}$ and the $W_{k}$ commute. \\
(vi) If the sequence $\left(\prod_{k=1}^{N}j_k(W_{\infty}^{-1}W_{k})\right)_{N\in\mathbb{N}^*}$
is norm bounded, the weak limit \eqref{lim-x[0,n]0} exists on $ \mathcal{H}_{\psi} $.\\
(v) Define, for $I\subset_{fin}\mathbb{N}^*$
\begin{equation}\label{df-xn-fi}
x_{I} := \prod_{k\in I}j_k(W_{\infty}^{-1}W_{k})
\end{equation}
If the series
\begin{equation}\label{norm-conv-cond}
\sum_{k=1}^{\infty}\left\|W_{\infty}^{-1}W_{k} - 1 \right\|
\end{equation}
converges, then the sequence $(x_{[1,N]})$ is Cauchy in norm.
\end{theorem}
\noindent\textbf{Proof}.
(i) follows from the fact that $\psi$ is exchangeable by construction and, for any
$g\in\mathcal{S}_{\infty}$ and $a\in\mathcal{A}$, one has
$$
\hat{\psi}(g\pi_{\psi}(a)) = \hat{\psi}(\pi_{\psi}(ga))
= \langle \Psi \, , \, \pi_{\psi}(ga)\Psi\rangle =\psi(ga) =\psi(a)
= \hat{\psi}(\pi_{\psi}(a))
$$
(ii) Notice that the map $I\subset_{fin}\mathbb{N}^*\mapsto x_{I}$, defined by \eqref{df-xn-fi}, is
a multiplicative functional, i.e.
$I\cap J =\emptyset \Rightarrow x_{I\cup J}=x_{I}x_{J}=x_{J}x_{I}$.\\
For any $N\in\mathbb{N}^*$, $g\in\mathcal{S}_{N}$ and $a_{[1,N]}\in\mathcal{A}_{[1,N]}$, one has
$$
\varphi (g(a))
\hbox{Tr}_{[1,N]}\left(\prod_{k=1}^{N} j_{k}(W_{k})g(a)\right)
\hbox{Tr}_{[1,N]}\left(\prod_{k=1}^{N}j_{k}(W_{\infty})
\prod_{k=1}^{N}j_{k}(W_{\infty}^{-1} W_{k}g(a))\right)
$$
$$
=\psi(x_{[1,N]}g(a))
\overset{\psi\circ g =\psi}{=}
\psi\left(g^{-1}\left(\left(x_{[1,N]}\right)g\left(a\right)\right)\right)
=\psi\left(g^{-1}\left(x_{[1,N]}\right)a\right)\\
$$
$$
=\psi\left(x_{[1,N]}x_{[1,N]}^{-1}g^{-1}\left(x_{[1,N]}\right)a\right)\\
=\varphi\left(x_{[1,N]}^{-1}g^{-1}\left(x_{[1,N]}\right)a\right)
$$
Since $a$ and $N $ are arbitrary, this proves that $\varphi$ is
$\mathcal{S}_{\infty}$--quasi invariant with cocycle
given by \eqref{cocycle-expl-prod-cse}. If $W_\infty$ commutes with all the $W_n$'s, \eqref{cocycle-expl-prod-cse} implies that $x_{g}$ is hermitean, i.e. $\varphi$ is
$\mathcal{S}_{\infty}$--strongly quasi--invariant.\\
To prove (iii) notice that, if $a_{[1,M]}, b_{[1,M]}\in\pi_{\psi}(\mathcal{A})''$ (one can always
suppose that $M$ is the same for both), then for any $M<N\in\mathbb{N}^*$, one has
\begin{equation}\label{lim-x[0,n]a}
\langle a_{[1,M]}\cdot\Psi \, , \,
\pi_{\psi}\left(\prod_{k=1}^{N}j_k(W_{\infty}^{-1}W_{k})\right)
b_{[1,M]}\cdot\Psi\rangle
\end{equation}
$$
= \langle a_{[1,M]}\cdot\Psi \, , \,
\pi_{\psi}\left(\prod_{k=1}^{M}j_k(W_{\infty}^{-1}W_{k})\right)
\pi_{\psi}\left(\prod_{k=M+1}^{N}j_k(W_{\infty}^{-1}W_{k})\right)
b_{[1,M]}\cdot\Psi\rangle
$$
$$
= \langle \Psi \, , \,
\pi_{\psi}\left(\prod_{k=M+1}^{N}j_k(W_{\infty}^{-1}W_{k})\right)
a_{[1,M]}^*\pi_{\psi}\left(\prod_{k=1}^{M}j_k(W_{\infty}^{-1}W_{k})\right)b_{[1,M]}\cdot\Psi\rangle
$$
$$
= \langle\Psi \, , \,
\pi_{\psi}\left(\prod_{k=M+1}^{N}j_k(W_{\infty}^{-1}W_{k})\right) \cdot\Psi\rangle \
\langle\Psi \, , \, a_{[1,M]}^*\pi_{\psi}\left(\prod_{k=1}^{M}j_k(W_{\infty}^{-1}W_{k})\right)
b_{[1,M]}\cdot\Psi\rangle
$$
where, in the last two equalities, we have used the fact that, for local algebras localized on
disjoint sets, both their commutativity and the factorizability of the state
$\langle\Psi \, ,  \, \cdot \, \Psi\rangle\big|_{\pi_{\psi}\left(\mathcal{A}\right)}$
are preserved under weak closures. So, one has
$$
\langle\Psi \, , \,
\pi_{\psi}\left(\prod_{k=M+1}^{N}j_k(W_{\infty}^{-1}W_{k})\right) \cdot\Psi\rangle \
\langle\Psi \, , \, a_{[1,M]}^*\pi_{\psi}\left(\prod_{k=1}^{M}j_k(W_{\infty}^{-1}W_{k})\right)
b_{[1,M]}\cdot\Psi\rangle
$$
$$
= \psi \left(\prod_{k=M+1}^{N}j_k(W_{\infty}^{-1}W_{k})\right)
\langle\Psi \, , \, a_{[1,M]}^*\pi_{\psi}\left(\prod_{k=1}^{M}j_k(W_{\infty}^{-1}W_{k})\right)
b_{[1,M]}\cdot\Psi\rangle
$$
\begin{eqnarray*}
&&= \hbox{Tr}_{\mathbb{N}^*}\left(\prod_{k=M+1}^{N}j_k(W_{\infty})\prod_{k=M+1}^{N}j_k(W_{\infty}^{-1}W_{k})\right)\\
&&\langle\Psi \, , \, a_{[1,M]}^*\pi_{\psi}\left(\prod_{k=1}^{M}j_k(W_{\infty}^{-1}W_{k})\right)
b_{[1,M]}\cdot\Psi\rangle
\end{eqnarray*}
$$
= \langle\Psi \, , \, a_{[1,M]}^*\pi_{\psi}\left(\prod_{k=1}^{M}j_k(W_{\infty}^{-1}W_{k})\right)
b_{[1,M]}\cdot\Psi\rangle
$$
because the $W_{k}$'s are density operators and $\hbox{Tr}_{\mathbb{N}^*}$ is factorizable in the given localization. This proves that the limit \eqref{lim-x[0,n]0} exists in the strongly finite
sense on $\pi_{\psi}(\mathcal{A}_{[1,M]})''\cdot\Psi$ and, since $M\in\mathbb{N}^*$ is arbitrary,
(i) follows.\\
(vi) follows from the fact that weak convergence of a norm bounded sequence of operators on a dense subspace of a Hilbert space implies weak convergence on the whole space.\\
To prove (v), we consider
\begin{equation}\label{SN-Cauchy1}
\| x_{[1,N]} - x_{[1,M]}\|
\overset{\eqref{df-xn-fi}}{=} \left\|\prod_{k=1}^{N} W_{\infty}^{-1}W_{k} - \prod_{k=1}^{M} W_{\infty}^{-1}W_{k}\right\|
\end{equation}
$$
\le  \left\|\prod_{k=1}^{M} W_{\infty}^{-1}W_{k}\right\|
 \ \left\|\prod_{k=M+1}^{N} W_{\infty}^{-1}W_{k} - 1\right\|
$$
Now we use the following identity, valid on any multiplicative semi--group $S$ with
distinguished element $1$: for any $M\le N\in\mathbb{N}^*$ and any sequence $(a_{h})_{h=M}^{N}$,
in $S$, one has
\begin{equation}\label{prod-id}
\prod_{h=M}^{N}a_{h} - 1
=\sum_{h=M}^{N}\left(\prod_{j=M}^{h-1}a_{j}\right)(a_{h} - 1)
\end{equation}
with the convention that $\prod_{j=M}^{M-1}a_{j}:=1$. The identity \eqref{prod-id} is proved by
induction. It clearly holds for $M=N$. Supposing that it holds for\\ $N-1>M$, one has
$$
\prod_{h=M}^{N}a_{h} - 1
=\prod_{j=M}^{N-1}a_{j}(a_{N}-1) + \prod_{j=M}^{N-1}a_{j} -1
\overset{induction}{=}\prod_{j=M}^{N-1}a_{j}(a_{N}-1)
+ \sum_{h=M}^{N-1}\left(\prod_{j=M}^{h-1}a_{j}\right)(a_{h} - 1)
$$
$$
=\sum_{h=M}^{N}\left(\prod_{j=M}^{h-1}a_{j}\right)(a_{h} - 1)
$$
Therefore, by induction, \eqref{prod-id} holds for any $N\in\mathbb{N}$.
Using \eqref{prod-id}, one obtains
\begin{equation}\label{SN-Cauchy2}
\| x_{[1,N]} - x_{[1,M]}\|\le  \left\|\prod_{k=1}^{M} W_{\infty}^{-1}W_{k}\right\|
 \ \left\|\prod_{k=M+1}^{N} W_{\infty}^{-1}W_{k} - 1\right\|
\end{equation}
$$
=\left\|\prod_{k=1}^{M} W_{\infty}^{-1}W_{k}\right\| \
\left\|\sum_{k=M+1}^{N}\left(\prod_{j=M+1}^{k-1}W_{\infty}^{-1}W_{j}\right)(W_{\infty}^{-1}W_{k} - 1) \right\|
$$
$$
\le\left\|\prod_{k=1}^{M} W_{\infty}^{-1}W_{k}\right\| \
\sum_{k=M+1}^{N}\left\|\left(\prod_{j=M+1}^{k-1}W_{\infty}^{-1}W_{j}\right)(W_{\infty}^{-1}W_{k} - 1) \right\|
$$
$$
\le\left\|\prod_{k=1}^{M} W_{\infty}^{-1}W_{k}\right\| \
\sum_{k=M+1}^{N}\left\|\prod_{j=M+1}^{k-1}W_{\infty}^{-1}W_{j}\right\|
\left\|W_{\infty}^{-1}W_{k} - 1 \right\|
$$
$$
\le\left\|\prod_{k=1}^{M} W_{\infty}^{-1}W_{k}\right\| \
\left(\sup_{k\ge M+2}\left\|\prod_{j=M+1}^{k-1}W_{\infty}^{-1}W_{j}\right\|\right)
\sum_{k=M+1}^{N}\left\|W_{\infty}^{-1}W_{k} - 1 \right\|
$$
$$
\le\prod_{k=1}^{M} \left\|W_{\infty}^{-1}W_{k}\right\| \
\left(\sup_{k\ge M+2}\prod_{j=M+1}^{k-1}\left\|W_{\infty}^{-1}W_{j}\right\|\right)
\sum_{k=M+1}^{N}\left\|W_{\infty}^{-1}W_{k} - 1 \right\|
$$
By assumption, the series \eqref{norm-conv-cond} converges. Hence, due to the inequality
$| \, \|W_{\infty}^{-1}W_{k}\| - 1 \, |\le \left\|W_{\infty}^{-1}W_{k} - 1 \right\|$, condition
\eqref{norm-conv-cond} implies that
\begin{equation}\label{norm-conv-cond-impl}
\sum_{k=1}^{\infty}| \, \|W_{\infty}^{-1}W_{k}\| - 1 \, |  < +\infty
\end{equation}
and, since by assumption $\|W_{\infty}^{-1}W_{k}\| \ne 0$ for each $k\in\mathbb{N}^*$,
it is known that condition \eqref{norm-conv-cond-impl} is necessary and sufficient for the
existence of a number $P>0$ such that:
\begin{equation}\label{exist-inf-prod}
\lim_{M\to+\infty}\prod_{k=1}^{M} \, \|W_{\infty}^{-1}W_{k}\| =P
\end{equation}
Let us prove that, if condition \eqref{norm-conv-cond} is satisfied, for $M$ large enough,
the \textit{sup} in the right hand side of \eqref{SN-Cauchy2} can be bounded by a number
independent of $N$. \eqref{exist-inf-prod} implies that, for any fixed $M$ and $N\ge M+1$,
\begin{equation}\label{prod-parz}
\prod_{k=M+1}^{N} \, \|W_{\infty}^{-1}W_{k}\|
= \frac{\prod_{k=1}^{N} \, \|W_{\infty}^{-1}W_{k}\|}{\prod_{k=1}^{M} \, \|W_{\infty}^{-1}W_{k}\|}
\overset{as \ N\to+\infty}{\longrightarrow}
\frac{P}{\prod_{k=1}^{M} \, \|W_{\infty}^{-1}W_{k}\|}>0
\end{equation}
It is convenient to introduce the notations
$$
P_{M]} := \prod_{k=1}^{M} \, \|W_{\infty}^{-1}W_{k}\| \quad;\quad
P_{(M,N]} := \prod_{k=M+1}^{N} \, \|W_{\infty}^{-1}W_{k}\|
$$
From \eqref{exist-inf-prod}, we know that, for any $\varepsilon\in (0,1)$ there exists $M_{\varepsilon}$ such that, for $M\ge M_{\varepsilon}$, and for any $N\ge M+1$,
\begin{equation}\label{SN-Cauchy3}
|P-P_{M]}|<\varepsilon \ ; \
\sum_{k=M+1}^{N}\left\|W_{\infty}^{-1}W_{k} - 1 \right\|
<\varepsilon\left(2\|P\| \left(\varepsilon + \frac{P}{P - \varepsilon}\right)\right)^{-1}
\end{equation}
Similarly, \eqref{prod-parz} implies that there exists $N_{M,\varepsilon}$ such that, for
$N\ge N_{M,\varepsilon}$,
$$
\left|P_{(M,\left.N\right]} - \frac{P}{P_{\left.M\right]}} \right|<\varepsilon
$$
Therefore
$$
|P_{(M,\left.N\right]}|
\le \left|P_{(M,\left.N\right]} - \frac{P}{P_{\left.M\right]}} \right| + \frac{P}{P_{\left.M\right]}}
<\varepsilon + \frac{P}{P - (P-P_{\left.M\right]})}
$$
On the other hand, since
$$
P - (P-P_{\left.M\right]})\ge P - |(P-P_{\left.M\right]})| > P-\varepsilon
$$
one has
\begin{equation}\label{est-prod-parz}
\sup_{M\ge M_{\varepsilon} \, , \, N\ge N_{\varepsilon} }|P_{(M,\left.N\right]}|
<\varepsilon + \frac{P}{P - \varepsilon}
\end{equation}
So that $M\ge M_{\varepsilon}$ and $N\ge M+2$,
$$
\| x_{[1,N]} - x_{[1,M]}\|  \, \overset{\eqref{SN-Cauchy2}}{\le} \,
\prod_{k=1}^{M} \left\|W_{\infty}^{-1}W_{k}\right\| \
\left(\sup_{k\ge M+2}\prod_{j=M+1}^{k-1}\left\|W_{\infty}^{-1}W_{j}\right\|\right)
\sum_{k=M+1}^{N}\left\|W_{\infty}^{-1}W_{k} - 1 \right\|
$$
$$
\overset{\eqref{exist-inf-prod} \, , \, \eqref{est-prod-parz}}{\le} \
2\|P\| \left(\varepsilon + \frac{P}{P - \varepsilon}\right)
\sum_{k=M+1}^{N}\left\|W_{\infty}^{-1}W_{k} - 1 \right\|
\overset{\eqref{SN-Cauchy3}}{\le} \ \varepsilon
$$
i.e. the sequence $(x_{[1,N]})$ is Cauchy in norm.
$\qquad\square$
\smallskip\\
\noindent\textbf{Remark}.
The identity \eqref{varphi-hat-psi} is the first example of a decomposition of the form
\eqref{struc-q-inv-st-cpct} for a locally compact but non--compact group.

\subsection{Markov Chains with commuting conditional density amplitudes }\label{sec:QMC-comm-CDA}

Recall that, for any exchangeable state $\psi$ on $\mathcal{A}$, all local sub--algebras of
$\mathcal{A}=\bigotimes_{n\in\mathbb{N}^*}\mathcal{B}$ are $\psi$--expected.
For $I\subseteq \mathbb{N}^*$, the sub--algebra of $\mathcal{A}$ localized in $I$ is denoted $\mathcal{A}_{I}$.
In this subsection, we fix an exchangeable state $\psi$ on $\mathcal{A}$, we denote
$E^{\psi}_{I}$ the (unique) Umegaki conditional expectation from $\mathcal{A}$ onto
$\mathcal{A}_{I}$ satisfying
$$
\psi\circ E^{\psi}_{I} = \psi  \quad (\mathcal{A}_{I}\hbox{ $\psi$--expected })
$$
Recall that, in the notation \eqref{notat-jn}, for $n\in \mathbb{N}^*$,
$j_{[n,n+1]}:=j_n\otimes j_{n+1}$, one has $j_{[n,n+1]}(\mathcal{B}\otimes \mathcal B) = \mathcal{A}_{[n,n+1]}$ and that
an $E^{\psi}_{I}$--\textbf{conditionally density amplitude (CDA)} localized in
$\mathcal{A}_{[n,n+1]}$ is an operator
$K_{n,n+1}\in\mathcal{B}\otimes \mathcal{B}$ satisfying
\begin{equation}\label{df-psi-CDA}
E^{\psi}_{\{n\}}(j_{[n,n+1]}(K_{n,n+1}^*K_{n,n+1}))
=E^{\psi}_{n]}\left(j_{[n,n+1]}(K_{n,n+1}^*K_{n,n+1})\right)
= 1_{\mathcal{A}_{n}}
\end{equation}
and that the family $(j_{[n,n+1]}(K_{n,n+1}))_{n\in \mathbb{N}^*}$ of CDA is
called \textbf{commutative} if the $C^*$--algebra generated by it is commutative.
With this notation, for any $g\in\mathcal S_\infty$ and any $a,b\in\mathcal{B}$,
one has
$$
gj_{[n,n+1]}(a\otimes b):=\begin{cases}
j_{[gn,g(n+1)]}(a\otimes b), & \mbox{if } g(n+1)>gn \\
j_{g(n+1)}(a)\otimes j_{gn}(b), & \mbox{if } gn>g(n+1)
                          \end{cases}
$$

\begin{remark}{\rm
In the following proposition we use the formal language introduced in the remark after
Proposition \ref{prop:ex1-prod-sts}. More explicitly, extending to infinite sets the notion of
\textit{right--product}
\begin{equation}\label{df-ord-prod}
K_{[1,n]}:=\overset{\rightarrow}{\prod}_{n\in [1,N]} j_{[n,n+1]}(K_{n,n+1})
\end{equation}
$$
:= j_{[1,2]}(K_{1,2})j_{[2,3]}(K_{2,3})\cdots j_{[N,N+1]}(K_{N,N+1})
$$
(and similarly for \textit{left--product} $\overset{\leftarrow}{\prod}_{n\in [1,N]}$)
one obtains products like
\begin{equation}\label{df-ord-prod-inf}
K_{\mathbb{N}^*}:=\overset{\rightarrow}{\prod}_{n\in\mathbb{N}^*} j_{[n,n+1]}(K_{n,n+1})
\end{equation}
$$
:= j_{[1,2]}(K_{1,2})j_{[2,3]}(K_{2,3})\cdots j_{[n,n+1]}(K_{n,n+1}) \cdots
$$
which are formal expressions whose precise meaning is given by the general theory of quantum
Markov chains (see for example \cite{[AcElGLuSou2024b]} for more details).
On the other hand, the use of these formal expressions clearly indicates that, dealing with quantum
Markov chains, and without the assumption that the CDAs are \textbf{commutative},
the natural cocycles have the form
$\varphi(g(a)) = \varphi(y^*_{g}ay_{g})$ and not $\varphi(g(a)) = \varphi(x_{g}a)$, so that
the natural notion of quasi--invariant state is more general than the one discussed
in the present paper.
This more general notion will be discussed in a paper in preparation.
}
\end{remark}

\begin{theorem}\label{thm:QI-MC}\upshape
In the above notations, given a sequence $K_{n,n+1}\in\mathcal{B}\otimes \mathcal{B}$ of \textbf{invertible} CDAs and an exchangeable state $\psi$ on $\mathcal{A}$, the state
$\varphi$ on $\mathcal{A}$ defined, in the notation \eqref{df-ord-prod-inf}, by
\begin{equation}\label{QI-MC}
\varphi(a)
:=\psi\left(\left(\overset{\leftarrow}{\prod}_{n\in\mathbb{N}^*} j_{[n,n+1]}(K_{n,n+1}^{*})\right)a
\left(\overset{\rightarrow}{\prod}_{n\in\mathbb{N}^*} j_{[n,n+1]}(K_{n,n+1})\right)\right)
\end{equation}
is well defined and satisfies, $\forall N\in\mathbb{N}^*$, $\forall g\in\mathcal{S}_{\infty}$,
such that $\Lambda_g\subseteq [1,N]$ and $\forall a\in \mathcal{A}_{[1,N]}$,
\begin{equation}\label{varphi-gen-QI}
\varphi(g(a))
= \varphi\left(y_{[1,N];g}^* a y_{[1,N];g}\right)
\end{equation}
where
\begin{equation}\label{gen-cocy}
y_{[1,N];g} :=
\end{equation}
$$
\left(\overset{\rightarrow}{\prod}_{n\in [1,N]} j_{[g^{-1} n,g^{-1}(n+1)]}(K_{n,n+1})\right)
\left(\overset{\rightarrow}{\prod} _{n\in [1,N]} j_{[n,n+1]}(K_{n,n+1})\right)^{-1}
\in\mathcal{A}_{[1,N]}
$$
In particular, if all the $K_{n,n+1}$ are in the centralizer of $\psi$, then $\varphi$
is a $\mathcal{S}_{\infty}$--strongly quasi--invariant state with cocycle
\begin{equation}\label{xg-MC-centr-cocy}
x_g=
|(y_{[1,N];g}^*|^2
= \left|\left(\overset{\leftarrow}{\prod} _{n\in [1,N]} j_{[n,n+1]}(K^*_{n,n+1})\right)^{-1}
\left(\overset{\leftarrow}{\prod}_{n\in [1,N]} j_{[g^{-1} n,g^{-1}(n+1)]}(K^*_{n,n+1})\right)
\right|^2
\end{equation}
If the $K_{n,n+1}$ are in the centralizer of $\psi$ and mutually commute, then,
$\forall N\in\mathbb{N}^*$, $\forall g\in\mathcal{S}_{\infty}$,
such that $\Lambda_g\subseteq [1,N]$,  the cocycle \eqref{xg-MC-centr-cocy} takes the form
\begin{equation}\label{xg-MC-centr-cocy-comm}
x_g=\prod_{n\in [1,N]}j_{[n,n+1]}(|K_{n,n+1}|^2)^{-1}) j_{[g^{-1}n,g^{-1}(n+1)]}(|K_{n,n+1}|^2)
,\;\;\;\forall g\in\mathcal{S}_{\infty}
\end{equation}
\end{theorem}
\noindent\textbf{Proof}.
The first statement follows from the commutativity of local algebras localized
on disjoint sets of $\mathbb{N}^*$ and from \eqref{df-psi-CDA}.
In fact, if $g\in\mathcal{S}_{\infty}$, $N\in\mathbb{N}^*$ and $a\in\mathcal{A}_{[1,N]}$, one
can always suppose that $\Lambda_g\subseteq [1,N]$. Under this condition, one has:
$$
\varphi(a)
\overset{\eqref{QI-MC}}{=}
\psi\left(\left(\overset{\leftarrow}{\prod} _{n\in\mathbb{N}^*} j_{[n,n+1]}(K_{n,n+1}^{*})\right)a
\left(\overset{\rightarrow}{\prod}_{n\in\mathbb{N}^*} j_{[n,n+1]}(K_{n,n+1})\right)\right)
$$
$$
= \psi\left(\left(\overset{\leftarrow}{\prod} _{n\in [1,N]^c} j_{[n,n+1]}(K_{n,n+1}^{*})\right)
\left(\overset{\leftarrow}{\prod} _{n\in [1,N]} j_{[n,n+1]}(K_{n,n+1}^*)\right)a\right.
$$
$$
\left.\left(\overset{\rightarrow}{\prod}_{n\in [1,N]} j_{[n,n+1]}(K_{n,n+1})\right)
\left(\overset{\rightarrow}{\prod}_{n\in [1,N]^c} j_{[n,n+1]}(K_{n,n+1})\right) \right)\\
$$
\begin{equation}\label{fi-a-well-def}
\overset{\eqref{df-psi-CDA}}{=} \psi\left(
\left(\overset{\leftarrow}{\prod} _{n\in [1,N]} j_{[n,n+1]}(K_{n,n+1}^*)\right)a
\left(\overset{\leftarrow}{\prod} _{n\in [1,N]} j_{[n,n+1]}(K_{n,n+1})\right)\right)
\end{equation}
So the right hand side of \eqref{QI-MC} is well defined on all local elements of $\mathcal{A}$.
With the same notations, and using the fact that $g\Lambda_g=\Lambda_g\subseteq [1,N]$,
\eqref{varphi-gen-QI} follows from

$$
\varphi(g(a))
\overset{\eqref{QI-MC}}{=}
\psi\left(\left(\overset{\leftarrow}{\prod} _{n\in\mathbb{N}^*} j_{[n,n+1]}(K_{n,n+1}^{*})\right)g(a)
\left(\overset{\rightarrow}{\prod}_{n\in\mathbb{N}^*} j_{[n,n+1]}(K_{n,n+1})\right)\right)
$$
$$
= \psi\left(g\left(\left(\overset{\leftarrow}{\prod} _{n\in\mathbb{N}^*} g^{-1}(j_{[n,n+1]}(K_{n,n+1}^*))\right)a
\left(\overset{\rightarrow}{\prod}_{n\in\mathbb{N}^*} g^{-1}\left(j_{[n,n+1]}(K_{n,n+1})\right)\right)\right)\right)\\
$$
$$
= \psi\left(\left(\overset{\leftarrow}{\prod} _{n\in\mathbb{N}^*} j_{[g^{-1}n,g^{-1}(n+1)]}(K_{n,n+1}^*)\right)a
\left(\overset{\rightarrow}{\prod}_{n\in\mathbb{N}^*} j_{[g^{-1} n,g^{-1}(n+1)]}\left(K_{n,n+1}\right)\right)\right)\\
$$
$$
= \psi\left(\left(\overset{\leftarrow}{\prod} _{n\in [1,N]^c} j_{[n,n+1]}(K_{n,n+1}^{*})\right)
\left(\overset{\leftarrow}{\prod}_{n\in [1,N]} j_{[g^{-1}n,g^{-1}(n+1)]}(K_{n,n+1}^*)\right)a\right.
$$
$$
\left.\left(\overset{\rightarrow}{\prod}_{n\in [1,N]} j_{[g^{-1} n,g^{-1}(n+1)]}(K_{n,n+1})\right)
\left(\overset{\rightarrow}{\prod}_{n\in [1,N]^c} j_{[n,n+1]}(K_{n,n+1})\right) \right)\\
$$
\begin{equation}\label{gen-cocy1}
\overset{\eqref{df-psi-CDA}}{=}
\psi\left(\left(\overset{\leftarrow}{\prod} _{n\in [1,N]} j_{[g^{-1}n,g^{-1}(n+1)]}(K_{n,n+1}^*)\right)a
\left(\overset{\rightarrow}{\prod}_{n\in [1,N]} j_{[g^{-1} n,g^{-1}(n+1)]}(K_{n,n+1})\right)\right)
\end{equation}
$$
=\psi\left(
\left(\overset{\leftarrow}{\prod} _{n\in [1,N]} j_{[n,n+1]}(K_{n,n+1}^{*})\right)
\right.
$$
$$
\left.
\left(\overset{\leftarrow}{\prod} _{n\in [1,N]} j_{[n,n+1]}(K_{n,n+1}^{*})\right)^{-1}
\left(\overset{\leftarrow}{\prod} _{n\in [1,N]}j_{[g^{-1}n,g^{-1}(n+1)]}(K_{n,n+1}^*)\right)
\right.
a
$$
$$
\left.
\left(\overset{\rightarrow}{\prod}_{n\in [1,N]} j_{[g^{-1} n,g^{-1}(n+1)]}(K_{n,n+1})\right)
\left(\overset{\rightarrow}{\prod} _{n\in [1,N]} j_{[n,n+1]}(K_{n,n+1})\right)^{-1}
\right.
$$
\begin{equation}\label{gen-cocy2}
\left.
\left(\overset{\rightarrow}{\prod} _{n\in [1,N]} j_{[n,n+1]}(K_{n,n+1})\right)\right)
\end{equation}
using \eqref{gen-cocy}, \eqref{gen-cocy2} becomes
$$
\varphi(g(a))=\psi\left(
\left(\overset{\leftarrow}{\prod} _{n\in [1,N]} j_{[n,n+1]}(K_{n,n+1}^{*})\right)
y_{[1,N];g}^* a y_{[1,N];g}
\left(\overset{\rightarrow}{\prod} _{n\in [1,N]} j_{[n,n+1]}(K_{n,n+1})\right)\right)
$$
$$
=\psi\left(
\left(\overset{\leftarrow}{\prod} _{n\in\mathbb{N}^*} j_{[n,n+1]}(K_{n,n+1}^{*})\right)
y_{[1,N];g}^* a y_{[1,N];g}
\left(\overset{\rightarrow}{\prod} _{n\in\mathbb{N}^*} j_{[n,n+1]}(K_{n,n+1})\right)\right)
$$
$$
=\varphi\left(y_{[1,N];g}^* a y_{[1,N];g}\right)
$$
which, given \eqref{gen-cocy}, is \eqref{varphi-gen-QI}.
\eqref{xg-MC-centr-cocy} and \eqref{xg-MC-centr-cocy-comm} are simple consequences of \eqref{varphi-gen-QI}.
$\qquad\square$\\
\textbf{Remark}.
Theorem \ref{thm:QI-MC}, in particular \eqref{xg-MC-centr-cocy} and \eqref{xg-MC-centr-cocy-comm}, gives another example of locally trivial, in the sense of Definition \ref{df:loc-triv-cocy}, but in general non--trivial cocycle for the action of $\mathcal{S}_{\infty}$  on $\mathcal{A}$.

\section{Unitary representations associated to strongly quasi--invariant states}
\label{Unitary-reprs-str-qi-st}


Let $G\subseteq\hbox{Aut}(\mathcal{A})$ be a group of $*$--automorphisms of
$\mathcal{A}$. In the following the cyclic representation of $\{\mathcal{A},\varphi\}$
shall be denoted $\{\mathcal{H}_{\varphi},\pi,\Phi\}$.
\begin{theorem}\label{exixt-Un-rep-G-on-G-fi}{\rm
If $\varphi$ is $(G,x)$--strongly quasi--invariant,
there exists a unique unitary representation $U$ of $G$
on $\mathcal{H}_{\varphi}$ characterized by the property
\begin{equation}\label{df-Ug}
U_g\pi(a)\Phi=\pi(g(a)x^{1/2}_{g^{-1}}))\Phi\ ;\quad\forall\,a\in\mathcal{A}
\end{equation}
where $x^{1/2}_{g^{-1}}$ is the square root of $x_{g^{-1}}$. Moreover
\begin{equation}\label{*-auto-ug}
u_{g}(\pi(a)) := U^*_g\pi(a)U_g=\pi(g^{-1}(a))
\ ;\quad\forall\,g\in G\ ;\quad\forall\,a\in\mathcal{A}
\end{equation}
}\end{theorem}
\textbf{Proof\/}. In the above notations, for any $a,b\in\mathcal{A}$ and $g\in G$,
one has
\begin{eqnarray*}
\langle U_g\pi(a)\Phi,\ U_g\pi(b)\Phi\rangle
&\overset{\eqref{df-Ug}}{=}&
\langle\pi(g(a)x^{1/2}_{g^{-1}})\Phi,\pi(g(b)x^{1/2}_{g^{-1}})\Phi\rangle\\
&=&\langle \Phi,\pi(x^{1/2}_{g^{-1}}g(a^*b)x^{1/2}_{g^{-1}})\Phi\rangle\\
&=&\varphi(x^{1/2}_{g^{-1}}g(a^*b)x^{1/2}_{g^{-1}})\\
&=&\varphi(g(a^*b)x_{g^{-1}})\\
&=&\varphi(g(a^*b\cdot g^{-1}(x_{g^{-1}})))\\
&\overset{\eqref{df-G-q-inv}}{=}& \varphi(x_{g}a^*bg^{-1}(x_{g^{-1}}))\\
&=&\varphi(a^*bg^{-1}(x_{g^{-1}})x_{g})\\
&\overset{\eqref{xg-1}}{=}& \varphi(a^*b)\\
&=&\langle \pi(a)\Phi,\pi(b)\Phi\rangle
\end{eqnarray*}
Thus $U_g$ is isometric. Now, if $g,g'\in G$, then
\begin{eqnarray*}\label{right-hand-side}
U_gU_{g'}\pi(a)\Phi &=& U_g\pi(g'(a)x^{1/2}_{g'^{-1}})\Phi\\
 &=& \pi(g[g'(a)x^{1/2}_{g'^{-1}}]x^{1/2}_{g^{-1}})\Phi\\
 &=& \pi(gg'(a) g(x^{1/2}_{g'^{-1}})x^{1/2}_{g^{-1}})\Phi\\
&=& \pi(gg'(a) g(x^{1/2}_{g'^{-1}})x^{1/2}_{g^{-1}})\Phi\\
&\overset{\eqref{xs=g(ys)}}{=}&
\pi(gg'(a) g(x_{g'^{-1}})^{1/2}x^{1/2}_{g^{-1}})\Phi\\
&\overset{\eqref{xg-comm-xg'} \, , \, \eqref{xg-comm-g(xg')}}{=}&
\pi(gg'(a) (g(x_{g'^{-1}})x_{g^{-1}})^{1/2})\Phi\\
&\overset{\eqref{xg-comm-g(xg')}}{=}&
\pi(gg'(a) (x_{g^{-1}}g(x_{g'^{-1}}))^{1/2})\Phi\\
&\overset{\eqref{xg-1G-left-cocy}}{=}&
\pi(gg'(a) x_{g'^{-1}g^{-1}}^{1/2})\Phi\\
&=&\pi(gg'(a) x_{(gg')^{-1}}^{1/2})\Phi\\
&=&U_{gg'}\pi(a)\Phi
\end{eqnarray*}
In particular, putting first $g'=g^{-1}$ and then $g=g'^{-1}$, one finds\\
$U_gU_{g^{-1}} = U_{g^{-1}}U_g = U_{e} = 1$, so that
\begin{equation}\label{Ug*=Ug-1}
U_g^*=U_{g^{-1}}
\end{equation}
Thus $U$ is a unitary representation.

Finally, for $a,b \in\mathcal{A}$ and $g\in G$, one has
\begin{eqnarray*}\label{r-h-s}
U^*_g\pi(a)U_g\pi(b)\Phi &=& U_{g^{-1}}\pi(a\cdot g(b)\cdot x^{1/2}_{g^{-1}})\Phi\\
 &=& \pi(g^{-1}[ag(b)x^{1/2}_{g^{-1}}]\cdot x^{1/2}_g)\Phi\\
&=& \pi(g^{-1}(a)b\cdot[g^{-1}(x^{1/2}_{g^{-1}})x^{1/2}_g])\Phi
\end{eqnarray*}
and, by (\ref{x(-s)g}) with $s=1/2$, the right hand side of (\ref{r-h-s}) is equal to
$$
\pi(g^{-1}(a)b)\Phi=\pi(g^{-1}(a))\pi(b)\Phi
$$
which implies (\ref{*-auto-ug}) by the cyclicity of $\Phi$.
$\square$

\section{Structure of strongly quasi--invariant states}\label{sec:Struct-strongly-QIS}

In this section we assume that $\mathcal{A}$ is a von Neumann algebra acting on a Hilbert space $\mathcal H$, $G$ is a group of normal $*$--automorphisms of $\mathcal{A}$ such that the action $g\mapsto \tau_g$ on $\mathcal{A}$ is strongly continuous, $\varphi$ is a normal faithful $G$--strongly quasi--invariant on $\mathcal{A}$ and the map $g\in G\mapsto x_g$ is strongly continuous. Moreover, the commutative $C^*$--algebra generated by the $x_g$ will be denoted by $\mathcal{C}$. In the following if no confusion can arise $\tau_g$ will be simply denoted by $g$.

\subsection{ The case of compact $G$ }\label{sec:Struct-strong-QIS-cpt-cse}


In this subsection we assume that $G$ is a \textbf{compact} group of normal\\ $*$--automorphisms
of $\hbox{Aut}(\mathcal{A})$ with \textbf{normalized} Haar measure $dg$.
\begin{lemma}\label{Spect}{\rm
If $G$ is a compact group and the map $g\in G\mapsto x_g$ is strongly continuous, then $\{ x_g\}_{g\in G}$ is uniformly bounded. In particular, there exist $0<S_1<S_2$ such that
\begin{equation}\label{spec}
\mbox{Spec}(x_g)\subset[S_1,\;S_2]
\end{equation}
}
\end{lemma}
\textbf{Proof\/}. Since $G$ is compact and for all $\xi\in\mathcal H$, $g\mapsto x_g\xi$ is continuous, then $\mbox{Sup}_{g\in G}||x_g\xi||<\infty$. Therefore by Banach--Steinhaus Theorem $\{ x_g\}_{g\in G}$ is uniformly bounded. Finally \eqref{spec} follows from the fact that for all $g\in G$, $x_g$ is an invertible positive operator.
$\square$

\begin{theorem}\label{thm:compact-case}{\rm
In the above assumptions the element
\begin{equation}\label{struct-kappa-g}
\kappa:=\int_Gx_{g}dg    \in \mathcal{C}\subseteq\hbox{Centrz}(\varphi) \subseteq\mathcal{A}
\end{equation}
$\kappa$ is an invertible operator with bounded inverse (hence $\kappa^{-1}$ belongs
to $\mathcal{C}$).
}\end{theorem}
\textbf{Proof\/}. Notice that the $C^*$--algebra $\mathcal{C}$ is abelian and contains the identity of $\mathcal{A}$ ($1_{\mathcal{A}}=x_e$).
Therefore, by Gelfand Theorem ((see \cite{[Sakai71]}, section 1.1.9) $\mathcal{C}$ can be
identified to the algebra
\begin{equation}\label{df-mathcal-C}
\mathcal{C}_{\mathbb{C}}(\mathcal S) := \{\hbox{ continuous complex valued functions
on a compact space $\mathcal S$} \}
\end{equation}
Since $\mathcal S$ is compact, in this identification, the $x_{g}$, being positive
and invertible, become strictly positive functions on a compact set.
Using the functional realization of the algebra $\mathcal{C}$, we realize the $x'_gs$ as
continuous strictly positive functions on the compact Hausdorff space $\mathcal S$.
Then each function $s\in \mathcal S \mapsto x_{g}(s)$ is bounded away from zero and from Lemma \ref{Spect}
$$
\kappa(s)=\int_Gx_{g}(s)dg>0 \qquad,\qquad\forall s\in \mathcal S
$$
is continuous and bounded away from zero.
Hence $\kappa^{-1}$ is bounded\textcolor{red}{,} strictly positive and $\kappa^{-1}$ belongs
to $\mathcal{C}$.
$\square$
\begin{theorem}\label{thm:EG-G-comp}{\rm
Define
\begin{equation}\label{df-EG-G-comp}
E_G=\int_Ggdg
\end{equation}
Then
\begin{enumerate}
\item[(i)] For any $g\in G$, $E_G$ satisfies the following identity
\begin{equation}\label{EG-commutes-G-fix}
gE_G=E_G=E_Gg
\end{equation}
In particular
\begin{equation}\label{Range-EG-in-Fix(G)}
\hbox{Range}(E_G) = \hbox{Fix}(G):=\{a\in\mathcal{A}:\;\;\;g(a)=a
\textcolor{red}{ \, , \, } \forall g\in G\}
\end{equation}
and $E_G(\mathcal{A})$ is a $W^*$--sub--algebra of $\mathcal{A}$.
\item[(ii)] $E_G$ is a faithful Umegaki conditional expectation
from $\mathcal{A}$ onto\\ $E_G(\mathcal{A})=\hbox{Fix}(G)$.
\end{enumerate}
}\end{theorem}
\textbf{Proof}.
For each $a\in \mathcal{A}$
$$
\mathcal{A}\ni
(E_G(a))^*=\left(\int_Gg(a)dg\right)^* =\int_Gg(a)^*dg =\int_Gg(a^*)dg=E_G(a^*)
$$
Therefore $E_G(\mathcal{A})$ is a $*$--sub--space of $\mathcal{A}$. This proves (i).\\
Left--invariance of the Haar measure implies that, for each $a\in \mathcal{A}$ and $g\in G$,
$$
gE_G(a)=\int_G gh(a)dh =\int_G gh(a) d(gh) =\int_G g'(a)dg'
= E_G(a)
$$
This proves the first identity in \eqref{EG-commutes-G-fix}.
Similarly, right--invariance of the Haar measure of $G$
implies that, for each $a\in \mathcal{A}$ and $g\in G$,
$$
E_G(g(a)) = \int_G hg(a)dh = \int_G hg(a)dhg =\int_G g'(a)dg'= E_G(a)
$$
which is the second identity in \eqref{EG-commutes-G-fix}.
The first identity in \eqref{EG-commutes-G-fix} implies that
$\hbox{Range}(E_G)\subseteq \hbox{Fix}(G)$.
The converse inclusion is clear because, if $a\in \hbox{Fix}(G)$, then
$$
a=\int_{G}g(a)dg = E_G(a)
$$
$E_G(\mathcal{A})=\hbox{Fix}(G)$ which is a $W^*$--sub--algebra of $\mathcal{A}$.
This proves (i).\\
To prove (ii) notice that $E_G$ is a completely positive, identity preserving,\\ $*$--map
from $\mathcal{A}$ to $\mathcal{A}$ being a convex combination of automorphisms.\\
Moreover, for any  $a\in \mathcal{A}$, one has
$$
E_G^2(a) = \int_G  dg \, g(E_G(a))
\overset{\eqref{EG-commutes-G-fix}}{=} \int_G  dg \, E_G(a) = E_G(a)
$$
due to the normalization of the Haar measure of $G$.
So $E_G$ is a completely positive norm--$1$ projector on $\mathcal{A}$ and, by Tomijama's
theorem \cite{[Tomija57]}, $E_G(\mathcal{A})$ is a Umegaki conditional expectation from
$\mathcal{A}$ to $E_G(\mathcal{A})$.
Finally, if $a$ is an element of $\mathcal{A}$ such that $E_G(a^*a)=0$, then
$$
0 = \varphi(E_G(a*a)) = \varphi(\kappa a^*a) = \varphi(\kappa^{1/2} a^*a\kappa^{1/2})
$$
Since $\kappa$ is invertible and $\varphi$ is faithful, it follows that $a = 0$.
$\qquad\square$\\

\noindent The structure of strongly quasi--invariant faithful normal states
with respect to compact groups is described by the following theorem.
\begin{theorem}\label{struc-strong-quasi-invar}\upshape
Let $G$ be a \textbf{compact group} of normal $*$--automorphisms of $\mathcal{A}$ and let
$\varphi$ be a $G$--strongly quasi--invariant faithful normal state on $\mathcal{A}$.
Denote $x:g\in G\to x_{g}\in\mathcal{A}$ the (strongly continuous, hermitean) left--$G$--$1$--cocycle associated to $\varphi$ and let $\kappa$ be as in Theorem
\ref{thm:compact-case}, namely
\begin{equation}\label{df-kappa}
\kappa = \int_Gx_{g}dg
\end{equation}
Then $\varphi$ can be written in the form
\begin{equation}\label{struc-q-inv-st-cpct}
\varphi(a)=\left(\varphi\circ E_G\right)(\kappa^{-1}a)=:\varphi_G(\kappa^{-1}a)
 \quad ;\quad\forall\,a\in\mathcal{A}
\end{equation}
In particular $\varphi_G$ is a faithful $G$--invariant state and
\begin{equation}\label{constr--xg}
x_g=\kappa g^{-1}(\kappa^{-1})
\end{equation}
Moreover
\begin{equation}\label{EG(kappa-1)=1-cpct}
E_G(\kappa^{-1})=1
\end{equation}
\begin{equation}\label{kappa-kappa-g-1kappa-1-herm}
\kappa^{*} = \kappa \quad;\quad \kappa g^{-1}(\kappa^{-1})= g^{-1}(\kappa^{-1})\kappa
 \ , \ \forall g\in G
\end{equation}
Conversely, let $G$ and $\mathcal{A}$ be as in the statement of the theorem. Then, for any
$\kappa\in\mathcal{A}$, invertible with inverse in $\mathcal{A}$ and satisfying \eqref{EG(kappa-1)=1-cpct} and for any $G$--invariant faithful normal state $\varphi_{G}$
on $\mathcal{A}$, the state
$$
\varphi ( \, \cdot \, ) := \varphi_{G}\left(\kappa^{-1} \, \cdot \, \right)
$$
is a $G$--quasi--invariant faithful normal state on $\mathcal{A}$ with cocycle given by
\eqref{constr--xg}. If $\kappa$ also satisfies \eqref{kappa-kappa-g-1kappa-1-herm}, then
$\varphi ( \, \cdot \, )$ is also $G$--strongly quasi--invariant.
\end{theorem}
\textbf{Proof}.
If $\varphi$ is a $(G,x_.)$ strongly--quasi--invariant state, define
$\varphi_G:=\varphi\circ E_G$.
$\varphi_G$ is clearly a state and is faithful because $\varphi$ and $E_G$
are faithful. It is $G$--invariant because of \eqref{EG-commutes-G-fix}.
From Theorem \ref{thm:compact-case}, one knows that $\kappa$ is an invertible element of $\mathcal{A}$ with inverse in $\mathcal{A}$.
Moreover, for any $a\in\mathcal{A}$ one has:
$$
\varphi_G(\kappa^{-1}a)=\varphi(E_G(\kappa^{-1}a))
=\int_Gdg \varphi(g(\kappa^{-1}a))=\int_Gdg\varphi(x_{g}\kappa^{-1}a)
$$
$$
=\varphi(\kappa\kappa^{-1}a) =\varphi(a)
$$
which is \eqref{struc-q-inv-st-cpct}. In particular
$$
\varphi(x_{g}a) = \varphi(g(a))
\overset{\eqref{struc-q-inv-st-cpct}}{=} \varphi_G(\kappa^{-1}g(a))
=\varphi_G(g(\kappa^{-1})a)
$$
$$
=\varphi_G(\kappa^{-1}\kappa g(\kappa^{-1})a)
=\varphi(\kappa g(\kappa^{-1})a)
$$
and \eqref{constr--xg} follows because $\varphi$ is faithful.
$\varphi_G$ is also faithful because, if $0\ne a$ is a positive element in $\mathcal{A}$ such that
$\varphi_G(a)=0$, then
$$
0\le \varphi(a^{1/2}) \overset{\eqref{struc-q-inv-st-cpct}}{=} \varphi_G (\kappa^{-1}a^{1/2})
\le \varphi_G (\kappa^{-2})\varphi_G (a)=0
$$
against the assumption that $\varphi$ is faithful.
Since the Haar measure on a unimodular group (in particular on a compact group) is invariant by inversion, one gets
$$
E_G(\kappa^{-1})
=\int_Gg(\kappa^{-1})dg
\overset{\eqref{constr--xg}}{=} \int_G\kappa^{-1} x_{g^{-1}}dg
=\kappa^{-1} \int_Gx_{g^{-1}}dg
=\kappa^{-1}\kappa
=1
$$
which is \eqref{EG(kappa-1)=1-cpct}.\\
Conversely, that is $(G,x)$--quasi-invariant (strongly $(G,x)$--quasi-invariant) with cocycle
\eqref{constr--xg} is known from Theorem \ref{thm:triv-cocy}.
Faithfulness and normality follow from arguments similar to those used in the forst part of the
proof.
$\qquad\square$\smallskip\\
\textbf{Remark}.
If the algebra of fixed points of the action of $G$ on $\mathcal{A}$ is non--trivial, the decomposition \eqref{struc-q-inv-st-cpct}, i.e. $\varphi = \varphi_G(\kappa^{-1} \, \cdot \, )$,
is not unique. In fact, in this case there exists a positive invertible fixed point $k$ which is not
a multiple of the identity and such that $\varphi_G(k \, \cdot \, )$ is an invariant state on
$\mathcal{A}$. In this case
$$
\varphi
= \varphi_G(\kappa^{-1} \, \cdot \, )
= \varphi_G(k \ (\kappa k)^{-1} \ \cdot \, )
$$
We have seen, in Section \ref{sec:Expls-str-qis-perm}, that non uniqueness of the decomposition
\eqref{struc-q-inv-st-cpct} can still take place when the fixed point algebra of the action of $G$
is trivial, but its dual action has a convex set of invariant states of cardinality $>1$.

\subsection{Inductive limits of compact groups}\label{sec:Induct-lim-cpct-grps}

In this section, $(G_{N})_{N\in\mathbb{N}}$ will be an increasing sequence of compact sub--groups
of a group $G$ of normal $*$-automorphisms of a von Neumann algebra $\mathcal{A}$ such that

\begin{equation}\label{df-G-loc-comp}
G:=\bigcup_{N\in\mathbb{N}}G_{N}
\end{equation}
and, for $N\in\mathbb{N}$, we denote $d_{N}g$ is the Haar measure on $G_{N}$ and
\begin{equation}\label{df-EGN}
E_{G_{N}}=\int_{G_{N}}gd_{N} g
\end{equation}
\begin{proposition}\label{prop:project-(EG(N))}{\rm
The family $(E_{G_{N}})_{N\in\mathbb{N}}$ is a projective family of conditional expectations, i.e.
\begin{equation}\label{project-(EG(N))}
E_{G_{N+1}}E_{G_{N}}  = E_{G_{N+1}}   \quad,\quad\forall N\in\mathbb{N}
\end{equation}
}\end{proposition}
\textbf{Proof}.
Since $G_{N}\subseteq G_{N+1}$, one has
\begin{equation}\label{Rge(EG(N+1))-in-Rge(EG(N))}
\hbox{Range}(E_{G_{N+1}}) = \hbox{Fix}({G_{N+1}})\subseteq \hbox{Fix}({G_{N}})
= \hbox{Range}(E_{G_{N}})
\end{equation}
Therefore projectivity is equivalent to \eqref{project-(EG(N))}.
Moreover
$$
E_{G_{N+1}}E_{G_{N}} = E_{G_{N+1}}\int_{G_{N}}gd_{N} g = \int_{G_{N}}d_{N} g E_{G_{N+1}}g
\overset{\eqref{EG-commutes-G-fix}}{=} \int_{G_{N}}d_{N} g E_{G_{N+1}} = E_{G_{N+1}}
$$
$\qquad\square$
\begin{lemma}\label{df-G-quasi-inv-cond}{\rm
Suppose that a state $\varphi$ on $\mathcal{A}$ is $G$--quasi--invariant and
denote, for each $g\in G$, $g\in G\mapsto x_{g}$ the $G$--cocycle and
\begin{equation}\label{df-Ng-LC}
N_{g} := \min\{ N\in\mathbb{N} \colon g\in G_{N}\}
\end{equation}
Then, for each $N\in\mathbb{N}$, $\varphi$ is $G_{N}$--quasi--invariant with cocycle
\begin{equation}\label{xg-indep-N}
g\in G_{N}\mapsto x_{G_{N};g}
\end{equation}
satisfying
\begin{equation}\label{df-xg-LC}
x_{g} = x_{G_{N_{g}};g} \quad,\quad\forall  g\in G_N
\end{equation}
In particular, $\varphi$ is $G$--strongly quasi--invariant if and only if it is $G_{N}$--strongly quasi--invariant for each $N\in\mathbb{N}$.
}\end{lemma}
\textbf{Proof}.
Since any $g\in G$ belongs to some $G_{N}$, $N_{g}$ is well defined by \eqref{df-Ng-LC}.
Since, for each $N\in\mathbb{N}$, $G_{N}\subseteq G$, it is clear that, if $\varphi$
is $G$--quasi--invariant, it is also $G_{N}$--quasi--invariant. To prove that \eqref{df-xg-LC}
holds notice that, for any $N\in\mathbb{N}$ such that $g\in G_{N}$,
$$
\varphi(x_{g}a) = \varphi(g(a)) = \varphi(x_{G_{N};g}a)
$$
The faithfulness of $\varphi$ then implies that $x_{g}=x_{G_{N};g}$ for any $N\in\mathbb{N}$ such that $g\in G_N$.
Thus in particular \eqref{df-xg-LC} holds.\\
Conversely, suppose that $\varphi$ is $G_{N}$--quasi--invariant for each $N\in\mathbb{N}$
and notice that, since any $g\in G$ belongs to some $G_{N_0}$. Then one has, for any $N\ge N_{g}$,
$$
\varphi(x_{G_{N_{g}};g}a) = \varphi(g(a)) = \varphi(x_{g}a)
$$
The faithfulness of $\varphi$ then implies that, defining $x_{g}$ by the right hand side
of \eqref{df-xg-LC}, one has
$$
\varphi(g(a)) = \varphi(x_{g}a) \quad,\quad \forall g\in G
$$
i.e. that $\varphi$ is $G$--quasi--invariant with cocycle $g\in G\mapsto x_{g}$. This proves the
first statement of the lemma. Given this and \eqref{df-xg-LC}, the second statement is clear.
$\qquad\square$

\begin{proposition}\label{exists-auto-rep-G}{\rm
In the notations of section \ref{Unitary-reprs-str-qi-st},
the family $(u_{g})_{g\in G}$ of $*$--automorphisms of $\pi(\mathcal{A})$ defined by
\eqref{*-auto-ug} extends uniquely to a representation of $G$ into the normal (or equivalently
strongly continuous) $*$--automorphisms of $\pi(\mathcal{A})''$, still denoted with
the same symbol.
Denoting for each $N\in \mathbb{N}$,
\begin{equation}\label{df-hat-EGN}
\overline{E}_{G_{N}} := \int_{G_{N}}u_{g}d_{N}g
\end{equation}
the family $(\overline{E}_{G_{N}})$ is a projective family of normal Umegaki
conditional expectations. Moreover, for each $N\in \mathbb{N}$ and with
$E_{G_{N}}$ defined by \eqref{df-EGN}, one has
\begin{equation}\label{expl-form-hatEGN}
\overline{E}_{G_{N}}(\pi(a)) = \pi(E_{G_{N}}(a))
\quad,\quad\forall a\in\mathcal{A}
\end{equation}
Each $\overline{E}_{G_N}$ can be extended by continuity to a Umegaki conditional expectation,
denoted with the same symbol $\overline{E}_{G_N}$, onto the weak closure of
\begin{equation}\label{Range-hatEGN=Fix(uG)}
\hbox{Fix}(u(G_{N}))
:= \hbox{weak closure of }\hbox{Fix}(u(G_{N}))\subset \pi(\mathcal{A})''
\end{equation}
The family $(\overline{E}_{G_N})_{N\in\mathbb N}$ is a projective decreasing family of
normal Umegaki conditional expectations.
}\end{proposition}
\textbf{Proof}.
Each $\overline{E}_{G_N}$ is a normal map because each $u_{g}$ ($g\in G_{N}$) is normal and this
property is preserved under integration over a compact set.
By normality, it can be extended by continuity to $\pi(\mathcal{A})''$.
\eqref{expl-form-hatEGN} holds because
$$
\overline{E}_{G_{N}}(\pi(a))
=\int_{G_{N}}d_{G_{N}}g u_{g}(\pi(a))
=\int_{G_{N}}d_{G_{N}}g U_{g}^*(\pi(a))U_{g}
\overset{\eqref{*-auto-ug}}{=} \int_{G_{N}}d_{G_{N}} (\pi(g^{-1}a))
$$
$$
=\int_{G_{N}}d_{G_{N}} (\pi(ga))
= \pi\left(\int_{G_{N}}d_{G_{N}}ga\right)
= \pi(E_{G_{N}}(a))
$$
Therefore $\overline{E}_{G_{N}}$ is a Umegaki conditional expectation from $\pi(\mathcal{A})$
onto $\hbox{Fix}(G_{N})$. This implies that its extension is a Umegaki conditional
expectation from $\pi(\mathcal{A})''$ onto $\hbox{Fix}(u(G_{N}))$.
Since $(E_{G_N})_{N\in\mathbb N}$ is a projective decreasing family, the same is true for
$(\overline{E}_{G_N})_{N\in\mathbb N}$.
$\qquad\square$

\section*{Acknowledgements}
The authors are grateful to Eric Ricard whose precious suggestions allowed them
to considerably improve the presentation of several results in this paper.
A. Dhahri is a member of GNAMPA-INdAM.

\end{document}